\documentclass[11pt,a4paper]{JHEP3}
\usepackage{amsmath, amssymb}
\usepackage{euscript}
\def\be{\begin{eqnarray}}
 \def\ee{\end{eqnarray}}
 \def\0{\nonumber}

\def\A{{\bf A}}
\def\B{{\bf B}}
\def\D{{\bf D}}
\def\C{{\bf C}}

\newcommand\CA{{\cal A}}
\newcommand\CC{{\cal C}}
\newcommand\EA{\EuScript{A}}

\newcommand\EG{\EuScript{G}}

\def\sin{{\rm sin}}
\def\tg{{\rm tan}}

\def\sinh{{\rm sinh}}
\def\cosh{{\rm cosh}}
\def\th{{\rm tanh}}
\def\coth{{\rm coth}}
\def\sin{{\rm sin}}
\def\tg{{\rm tan}}

\def\sinh{{\rm sinh}}
\def\cosh{{\rm cosh}}
\def\th{{\rm tanh}}
\def\coth{{\rm coth}}
\def\k{\kappa}
\def\det{{\rm det}}
\def\q{\bar q}
\def\p{\bar p}
\def\f{\buildrel \rightarrow \over{f}}

\def\u{\buildrel \rightarrow \over{u}}

\preprint{SISSA/34/2007/EP\\\tt hep-th/0706.1025}

\title{Ghost story. I. Wedge states in the oscillator formalism}

\author{ L.Bonora\\
International School for Advanced Studies (SISSA/ISAS)\\
Via Beirut 2--4, 34014 Trieste, Italy, and INFN, Sezione di
Trieste\\
E-mail:   \email{bonora@sissa.it},}

\author{C.Maccaferri\\
  Theoretische Natuurkunde, Vrije Universiteit Brussel,
Physique Th\'eorique et Math\'ematique, Universit\'e Libre de Bruxelles,
and
The International Solvay Institutes
Pleinlaan 2, B-1050 Brussels, Belgium\\
E-mail: \email{cmaccafe@vub.ac.be},}

\author{R.J.Scherer Santos\\
Centro Brasileiro de Pesquisas Fisicas (CBPF-MCT)-LAFEX\\
R. Dr. Xavier Sigaud, 150 - Urca - Rio de Janeiro - Brasil - 22290-180\\
E-mail: \email{scherer@cbpf.br},}

\author{D.D.Tolla\\
Center for Quantum SpaceTime (CQUEST), Sogang University\\
Shinsu-dong 1, Mapo-gu, Seoul, Korea\\
E-mail:  \email{tolla@sogang.ac.kr}}

\abstract{This paper is primarily devoted to the ghost wedge states in string
field theory formulated with the oscillator formalism. Our aim is to prove,
using such formalism, that the wedge states can be expressed as
$|n\rangle = {\rm exp}{\left[\frac {2-n}2 \left({\cal L}_0+
{\cal L}_0^\dagger\right)\right]}|0\rangle$, separately in the matter
and ghost sector. This relation is crucial for instance in the
proof of Schnabl's solution. We start from
the exponentials in the rhs and wish to prove that they take precisely
the form of wedge states. As a guideline we first re-demonstrate this relation
for the matter part. Then we turn to the ghosts. On the way we
face the problem of `diagonalizing' infinite
rectangular matrices. We manage to give a meaning to
such an operation and to prove that the eigenvalues we obtain satisfy the
recursion relations of the wedge states.}


\begin{document}

\maketitle

\section{Introduction}

The analytic solution of string field theory  \cite{Schnabl05} has been
found by means of the CFT language.
All the most recent developments in this field are also formulated by means
of the same powerful formalism
\cite{Okawa1,Ellwood:2006ba,RZ06,ORZ,Schnabl:2007az,KORZ,Fuchs2,Fuchs3,Fuchs0,Okawa2,Okawa3},
while the 'old' oscillator formalism \cite{Samu,CST,GJ1,GJ2,Ohta} has remained
in the shadow. This is in sharp contrast with the developments in vacuum string field
theory of a few years ago, when the two formalisms played a parallel role in the search
for solutions (but in that case the ghost sector of the theory was almost irrelevant).
One of the reasons for this asymmetry is certainly the extreme simplicity
some tools needed in order to prove the string field theory solution, take in
the conformal field theory language. The typical example in this sense is provided
by the wedge states, an essential tool in the construction of the solution, which
take an astonishingly simple form in the conformal language. The second reason is,
on the other hand, the almost unapproachably complicated form of the oscillator
formalism, especially in the ghost sector.

It is nevertheless clear that it would be highly desirable to derive Schnabl's
solution \cite{Schnabl05}
also in the oscillator formalism. On one hand it is disturbing that this
formalism,
which has served so well in many developments in string theory, seems to have become
suddenly unfit to deal with the most recent progress. On the other hand finding a more
algebraic way to formulate the analytic solution to string field theory may open new
prospects and suggest new solutions. For instance, in vacuum string
field theory the oscillator formalism opened the way to finding a ``complete" set of
orthonormal projectors \cite{RSZ3,tope},
a sort of canonical way to classify solutions.

The road map to an algebraic proof of Schnabl's solution has been laid down by Okawa in
\cite{Okawa1}. In order for this to be applicable to the oscillator approach
(apart from minor details) two main conditions should be fulfilled.
The first is the
representation of the wedge states as an exponential of ${\cal L}_0+{\cal L}_0^\dagger$
\be
|n\rangle = e^{\frac {2-n}2 \left({\cal L}_0+{\cal L}_0^\dagger\right)}
|0\rangle\label{nL0}
\ee
The second is a meaningful description of the star product of such states
as $c_1|0\rangle$ or of the wedge states themselves. This paper will deal with
the first problem.
We will show below that eq.(\ref{nL0}) can be given a meaning separately in the matter
sector and in the ghost sector, the latter being the crux of the problem. The oscillator
formalism proves to be, if not as simple as the conformal formalism, at least
as effective.

The first problem we meet in dealing with the ghost sector is the normal ordering.
We are familiar with what we call below the {\it conventional} normal ordering,
which is very handy in the analysis of the perturbative string spectrum.
This option is discussed in detail in section 8, but, even if it is correct, it does not
appear to be a convenient choice. We use instead the {\it natural} normal ordering, which
is required by the SL(2,R) invariant vacuum. This appears to be the right option.
The second problem is the use of asymmetric bases, that is we are obliged to use
matrices which act on two different bases on the left and on the right. This
complication can be dealt with first of all because the relevant matrices
commute and then because we can find the basis that diagonalizes them.
This allows us to reduce the relation (\ref{nL0}) to a relation
of eigenvalues and show that the relevant recursion relations are satisfied.

The paper is organized as follows. In section 2 we re-derive the equivalence (\ref{nL0})
for the matter sector. The problem has already been solved in the oscillator formalism,
but we redo it here as a guideline for the ghost sector, and also because the derivation
is partially new and simpler. Section 3 contains preliminary materials concerning the
ghost sector. Section 4 concerns the integration of what
we call the Kostelecky--Potting equations, \cite{KP}.
In section 5 we diagonalize the infinite matrices we need in our problem.
In section 6 we prove the recursion relation implied by eq.(\ref{nL0})
for the ghost sector.

In section 7, for completeness, we repeat the same oscillator analysis for the twisted
ghost sector, although twisted ghosts do not seem to be relevant to the proof of
Schnabl's analytic solution. In section 8 we discuss the already mentioned
conventional normal ordering option and illustrate its difficulties.
Section 9 contains some conclusions.
Several appendices contain auxiliary materials and samples of calculations that
are  needed in the course of the paper.

\section{Matter wedge states}

The exponential in the RHS of (\ref{nL0})factorizes in matter and ghost part, therefore
it must be possible to deal with the two sectors separately. Let us start with the matter
sector. Our problem is therefore to prove the representation of matter wedge states, i.e.
to prove that
\be
|n\rangle = e^{ - \frac {n-2}2({\cal L}_0 +
{\cal L}_0^\dagger)}|0\rangle
= {\cal N}_n\, e^{-\frac 12 a^\dagger S_n a^\dagger}|0\rangle \label{matterwedge}
\ee
where $S_n$ and ${\cal N}_n$ are the appropriate matrix and normalization factor
for the $n$--th matter wedge state
in the discrete oscillator basis \cite{Furu}. This problem has already been solved
recently by
Fuchs and Kroyter, \cite{Fuchs1,Fuchs2,FKM}. We repeat here in detail the
derivation because it is a
useful guide to the more complicated case of the ghost wedge states, but
also because our derivation, although inspired by Fuchs and Kroyter's one,
is rather different.

Let us start from the vacuum that appears in (\ref{matterwedge}). It is defined in
the usual way as $a_n|0\rangle=0$ for $n\geq 0$.
The matter Virasoro generators ($n>0$ and $\alpha_n = \sqrt{n} a_n$) are
\be
L_n^{(X)} &=&\alpha_0 \sqrt{n} \,a_n +\sum_{k=1}^\infty \sqrt{k(k+n)} \,
a_k^\dagger a_{n+k}+
\frac 12 \sum_{k=1}^{n-1} \sqrt{k(n-k)}\,a_k\,a_{n-k}\label{Ln}\\
L_0^{(X)}  &=& \frac 12 \, \alpha_0^2 + \sum_{k=1}^\infty \,k\,a_k^\dagger\,
a_k\label{L0}\\
L_{-n}^{(X)} &=& \alpha_0 \sqrt{n} \,a_n^\dagger +\sum_{k=1}^\infty \sqrt{k(k+n)}
 \,a_{n+k}^\dagger a_k+
\frac 12 \sum_{k=1}^{n-1} \sqrt{k(n-k)}\,a_{n-k}^\dagger
\,a_k^\dagger\label{L-n}
\ee
Since the zero mode does not come into play in the problem we are dealing
with and, on the other hand, it commutes with everything, we are at liberty to ignore
it altogether.

Now from the definition of ${\cal L}_0+{\cal L}_0^\dagger$ we have
\be
{\cal L}_0+{\cal L}_0^\dagger &=& 2L_0^{(X)} + \sum_{n=1}^\infty
\frac {2(-1)^{n+1}}
{4n^2-1} (L_{2n}^{(X)}+ L_{-2n}^{(X)})\0\\
&=& a^\dagger A a^\dagger + a\,B\,a +a^\dagger C\,a\label{L0+L0}
\ee
where the matrices $A,B,C$ are:
\be
A_{pq}=B_{pq}= \frac 12\sum_{n=1}^\infty \ell_{n}\, \sqrt{pq}\delta_{p+q,n}\label{A}
\ee
\be
C_{pq} =  \sum_{n\geq 1} \ell_n \,\sqrt{pq}\, (\delta_{p+n,q}+\delta_{q+n,p})
+ p \ell_0 \delta_{p,q}\label{C}
\ee
where
\be
\ell_{2n+1}=0, \quad\quad \ell_{2n}=
\frac {2(-1)^{n+1}}{4 n^2-1},\quad\quad n\geq 0,\quad\quad \ell_n=0, \quad n<0\label{elln}
\ee
These matrices are symmetric, moreover they vanish for odd $p+q$. Therefore
they commute with the twist matrix $\hat C$:
\be
\hat C\, A= A\,\hat C,\quad\quad \hat C\, C=C\,\hat C\0
\ee

\subsection{The KP equations}

The way to prove (\ref{matterwedge}) has been shown years ago by Kostelecky and
Potting, \cite{KP}. The idea is to factorize the exponential of (\ref{L0+L0}) as follows
\be
e^{t\left(a^\dagger A\,a^\dagger+ a^\dagger C \,a+a\,B\,a\right)} =
e^\eta\, e^{a^\dagger\alpha a^\dagger}
\,e^{a^\dagger\gamma\,a}\, e^{a\,\beta\,a}
\label{KPmatter}
\ee
where we have introduced an arbitrary parameter $t$. Therefore
$\alpha,\beta,\gamma$ and $\eta$ are to be understood as functions of $t$.
Now one differentiate both sides, commutes to the left and equates.
The result is
\be
&&A = \dot\alpha - \frac {d\, e^\gamma}{dt} e^{-\gamma}(\alpha+ \alpha^T)+
\frac 12 (\alpha+ \alpha^T) e^{-\gamma^T}(\dot \beta+\dot \beta^T)e^{-\gamma}
(\alpha+ \alpha^T)\label{C1}\\
&&B =  e^{-\gamma^T} \dot\beta e^{-\gamma}\label{C2}\\
&&C =  \frac {d\, e^\gamma}{dt} e^{-\gamma}- (\alpha+ \alpha^T)
e^{-\gamma^T}(\dot \beta+\dot \beta^T)e^{-\gamma}\label{C3}\\
&& 0=\dot \eta -{\rm Tr}(\alpha e^{-\gamma^T}(\dot \beta+\dot \beta^T)
e^{-\gamma})
\label{C4}
\ee
We will refer to these as the KP equations.
In our case we have $A=B=B^T$. So from (\ref{C2}) we get $\beta=\beta^T$.
Moreover we are interested only in the symmetric part of $\alpha$, so we
symmetrize (\ref{C1}). Calling the symmetric part with the same symbol
$\alpha$: $\frac 12(\alpha+\alpha^T) \to \alpha$, finally one gets
\be
&&\dot\alpha = A + \{C,\alpha\} + 4 \alpha B \alpha\label{C1'}\\
&&\dot \beta = e^{\gamma^T}\,B \,e^{\gamma}\label{C2'}\\
&& \frac {d\, e^\gamma}{dt} e^{-\gamma}=C +4\alpha B\label{C3'}\\
&&\dot \eta = 2\, {\rm Tr}(\alpha B)\label{C4'}
\ee
We are actually interested only in $\alpha$ and $\eta$, with the initial
condition $\alpha(0)=0$. If $CA=AC$, it is easy to integrate (\ref{C1'})
and obtain
\be
\alpha(t) = A \frac {{\rm sinh}\left(\sqrt{C^2-4A^2}\,t\right)}
{ \sqrt{C^2-4A^2}\, {\rm cosh} \left(\sqrt{C^2-4A^2}\,t\right)
-C\, {\rm sinh}\left(\sqrt{C^2-4A^2}\,t\right)}
\label{commcase}
\ee
where we have assumed that $C^2-4A^2$ is a positive operator.
It so happens that in our case $A$ and $C$ commute and $C^2-4A^2$ is indeed
positive, so that the solution to the KP equations
is precisely (\ref{commcase}).

\subsection{$A$ and $C$ commute}

Let us show that $A$ and $C$ commute. We have
\be
A_{pq}=B_{pq}=  \frac 12 \sqrt{pq}\, \ell_{p+q}
\label{Anew}
\ee
\be
C_{pq} =  \sqrt{pq}\, \ell_{|p-q|}\label{Cnew}
\ee
These matrices are symmetric and vanish for odd $p+q$.

The commutator between the two is
\be
(AC-CA)_{pq} =\frac{ \sqrt{pq}}2 \sum_{l=1}^\infty\, l\,
(\ell_{p+l}\, \ell_{|l-q|} -  \ell_{|p-l|} \,\ell_{q+l})\label{AC-CA'}
\ee
One can show that numerically this commutator vanishes for all $p$ and $q$,
but this can be shown also analytically.

To start with suppose that $p,q$ are both even. It follows that the summation
extend over
all even $l$ $\to 2l$. Therefore, in this case, after some algebra,
\be
(AC-CA)_{pq} = 4 \sqrt{pq} (-1)^{\frac {p+q}2}\Big(f(p,q)-f(q,p)\Big)
\label{comm1}
\ee
where
\be
f(p,q)&=&\sum_{l=1}^\infty \frac l{((p+2l)^2-1)((q-2l)^2-1)}\label{fpq}\\
&=&-\frac 1{2((p+q)^2-4)} \left(1- \frac 12 \frac {p-q}{p+q}
\left(\psi(\frac 12+\frac p2)-\psi(\frac 12-\frac q2)\right)\right)\0
\ee
$\psi$ is the dilogarithm function $\psi(z)=\frac d{dz} {\rm log}\Gamma(z)$.
One of its remarkable properties is that
\be
\psi(\frac 12 +z)-\psi(\frac 12 -z)= \pi\, {\rm Tan} (\pi z)\0
\ee
Using this one can easily prove (since $p,q$ are both even) that
$f(p,q)=f(q,p)$. Therefore (\ref{comm1}) vanishes identically.

Similarly, when $p,q$ are both odd, it follows that the summation extends
over odd $l$ $\to 2l+1$. Therefore, after some elementary algebra
\be
(AC-CA)_{pq} =- 2 \sqrt{pq} (-1)^{\frac {p+q}2}\Big(g(p,q)-g(q,p)\Big)
\label{comm2}
\ee
where
\be
g(p,q) &=& \sum_{l=0}^\infty (2l+1)
\frac 1 {(p+2l)(p+2l+2)(q-2l)(q-2l-2)}\label{gpq}\\
&=& \frac{2+p-q-2pq}{2pq((p+q)^2-4)} + \frac {p-q}{2(p+q)}\cdot
\frac{\psi(\frac p2)-\psi(-\frac q2)}{(p+q)^2-4}\0
\ee
It is not immediately evident, but using
\be
\psi(z)-\psi(-z)= -\pi\, {\rm Cot} (\pi z)-\frac 1z\0
\ee
one can prove that indeed
\be
g(p,q)=g(q,p)\0
\ee
The two functions (\ref{fpq},\ref{gpq}) become singular when $p=q=1$,
but from the initial definition (\ref{comm1}) it is evident that the
commutator in this special case is 0. In conclusion,
again, (\ref{comm2}) vanishes. Therefore $A$ and $C$ commute.

\subsection{Diagonalization of $K_1$}

To know more about $A$ and $C$ we can diagonalize them, which can be done by
finding a basis of common eigenvectors. To this purpose it is crucial to notice that
both $A$ and $C$ commute with the matrix $F$ representing the operator
$K_1=L_1+L_1^\dagger$. For $K_1$ can be written as
\be
K_1= a^\dagger F\,a\label{K1}
\ee
where
\be
F_{pq} = \sqrt{pq}\, (\delta_{p+1,q}+ \delta_{q+1,p})\label{F}
\ee
We have
\be
[{\cal L}_0+{\cal L}_0^\dagger, K_1] = -2 a^\dagger FA\, a^\dagger
+2 a FA\,a + a^\dagger [C,F]\,a\label{L0L0K1}
\ee
where we have utilized $B=A=A^T$ and $F=F^T$. It is elementary
to prove with the CFT language that the left hand side of
(\ref{L0L0K1}) vanishes.

Now $AF$ is not zero, but we should symmetrize it, because it appears
among two $a$'s or two $a^\dagger$'s. We get in fact
\be
(AF+FA)_{pq} = \frac 12 \sqrt{pq}\,\left( (p+q+2)\ell_{p+q+1} +
(p+q-2)\ell_{p+q-1}\right)=0\label{AF+FA}
\ee
The result is easily obtained by inserting the expression for $\ell_n$'s.

Likewise, for $p>q$ we get
\be
[C,F]_{pq} = \sqrt{pq} \left((q-p-2)\ell_{p-q+1} +(q-p+2)
\ell_{p-q-1}\right)=0\0
\ee
and a similar result for $q>p$.

The last equation means in particular that $C$ and $F$ must be simultaneously
diagonalizable. On the other hand, from eq.(\ref{AF+FA}) we get that $A$ and
$F$ anticommute. But $\{\hat C, F\}=0$. Therefore $[\tilde A, F]=0$, where
$\tilde A = \hat C\,A$. So also $\tilde A$ must be simultaneously
diagonalizable with $F$.

As it turns out, this problem has already been solved by Rastelli,
Sen and Zwiebach, \cite{RSZ1}, who determined the non degenerate
basis of eigenvectors of $K_1$. For the reader's convenience we
summarize the relevant results. $K_1$ represents the action of the
differential  operator \be {\cal K}_1\equiv (1+z^2)\frac
d{dz}\label{K1'} \ee in the $z$ plane. We can use this to find its
eigenvectors and its eigenvalues. Introduce the sequence
$v=\{v_n\}$ and define \be v\cdot a^\dagger =\sum_{n=1}^\infty \,
v_n \, a_n^\dagger\0 \ee Then, forgetting the zero mode $a_0$, we
get \be [K_1,v\cdot a^\dagger]= (F\,v)\cdot
a^\dagger\label{commK1v} \ee A related operator is \be (\tilde
F)_{nm} =\sqrt{\frac mn}\,F_{nm}= (n-1)\delta_{n,m+1} +(n+1)
\delta_{n+1,m} \ee Now define \be f_w(z) = \sum_{n=1}^\infty \,w_n
\,z^n,\quad\quad v_n= \sqrt{n}\, w_n \label{fw} \ee Then \be {\cal
K}_1 \, f_w(z) = f_{\tilde F\,w}(z) + w_{1}\label{K1action} \ee
Now it is easy to integrate \be (1+z^2) \frac {df(z)}{dz}=\k
\,f(z)\0 + w_{1}\ee with suitable boundary conditions, we get \be
f_{\k} (z) =- \frac 1{\k} \left(1- e^{\k \,
\arctan(z)}\right)\label{RSZfk} \ee Equating (\ref{fw}) to
(\ref{RSZfk}) we can extract the eigenfunctions \be w_n(\k) =
\frac 1{2 \pi i} \oint dz\, \frac {f_{\k}(z)}{z^{n+1}}\label{wn}
\ee The eigenfunctions found in this way, when suitably
normalized, form a complete orthonormal system.

\subsection{Diagonalization of $\Delta, \tilde A$ and $C$}

In the solution (\ref{commcase}) one of the building blocks is the
term $\Delta= C^2-4A^2= C^2 - 4 \tilde A^2$, the discriminant in the
integration of the
KP differential equation. It is possible to cast it in a very simple form.
Indeed one finds (see Appendix A)
\be
(C^2-4\,A^2)_{pq} &=&4\,\sqrt{pq} \, (-1)^{p+q}\,\frac{p+q}{2(p-q)((p-q)^2-4)} \0\\
&\cdot&
\left( \psi\left(\frac 12-\frac p2\right)- \psi\left(\frac 12+\frac p2\right)
+\psi\left(\frac 12+\frac q2\right) - \psi\left(\frac 12-\frac q2\right)\right)
\0\\
&=&4\,\sqrt{pq} \, (-1)^{p+q}\,\frac{p+q}{2(p-q)((p-q)^2-4)} \left(\pi\,
{\rm tan}(\pi q)-
\pi\, {\rm tan}(\pi p)\right) \0\\
&=& \frac 12 \pi^2\,\left( p^2\,\delta_{p,q}+\frac 14
\sqrt{pq} \,(p+q)( \delta_{p,q+2}+\delta_{p+2,q})\right)\label{C2-4A2}
\ee
Therefore $\Delta$ is a Jacobi matrix (it has only three nonvanishing diagonal
lines) and it is clearly positive. We know that, in the $v_n(\k)$ basis, this
matrix is diagonal, and in the form (\ref{C2-4A2}) it is rather easy
to find the corresponding eigenvalues.

Indeed let us use the representation (Fuchs and Kroyter) \be
v_n(\k) = {\cal N}(\k) \frac {i^{n-1} \sqrt{n}}{2\pi}
\,\int_{-\infty}^{\infty} du\, \frac {e^{-i\k u} {\rm
tanh}^{n-1}(u)}{{\rm cosh}^2(u)}\label{basis} \ee for the basis of
eigenvectors of $K_1$. Then \be &&\sum_{n=1}^\infty
(C^2-4A^2)_{pn}\,v_n(\k) = \frac {\pi^2}2 \sum_{n=1}^\infty \left(
p^2 \delta_{p,n} +\frac 14 \sqrt {pn} (p+n) (\delta_{p,n+2}+
\delta_{p+2,n})\right)\cdot \0\\
&&~~~~~~~~~~~~~~~~~~~~~~\cdot \frac{{\cal N}(\k)}{2\pi}
\int_{-\infty}^{\infty} du\, \frac {e^{-i\k u}}{{\rm cosh}^2(u)}
\, i^{n-1}
\sqrt{n}\, {\rm tanh}^{n-1}(u)\0\\
&=& \frac {\pi^2}2 \left( \frac{i^{p-1}\sqrt{p}{\cal N}(\k)}{2\pi}
 \int_{-\infty}^{\infty} du\, \frac {e^{-i\k u}}{{\rm cosh}^2(u)}\right.\cdot\0\\
&& \left. \cdot \left(p^2\, {\rm tanh}^{p-1}(u)-\frac 12 (p-1)(p-2)\,
{\rm tanh}^{p-3}(u) -\frac 12 (p+1)(p+2)\, {\rm tanh}^{p+1}(u)\right)\right)\0\\
&=& -\frac {\pi^2}4 \left( \frac{i^{p-1}\sqrt{p}\,{\cal N}(\k)}{2\pi}
 \int_{-\infty}^{\infty} du\, e^{-i\k u} \frac {\partial^2}{\partial u^2}
\left( \frac {{\rm tanh}^{p-1}(u)}{{\rm cosh}^2(u)}\right)\right)\0\\
&=& \frac {\pi^2}4 \k^2\, v_p(\k) \label{eigenvalueeq}
\ee
after integration by parts.

What remains for us to do is to derive the eigenvalues of $\tilde A$ and $C$.
As for the diagonalization of $\tilde A$, it turns out
that it has already been done by Rastelli, Sen and Zwiebach,
\cite{RSZ1}.
In fact their operator $B_{nm}$ defined in (5.18) coincides exactly with
\be
\tilde A_{pq}=\frac 12 \,(-1)^p\,\sqrt{pq} \,\ell_{p+q} =  \frac {\sqrt{pq} \,
(-1)^{\frac {q-p}2+1}}{(p+q)^2-1}\label{A=B}
\ee
with even $p+q$ and zero otherwise.
Therefore, using their result, the eigenvalue of $\tilde A$ is
\be
\tilde A(\k)=-\frac {\k\pi}{4 {\rm sinh}(\frac {\k\pi}2)}\label{A(k)}
\ee

Now, using (\ref{A(k)}), and $\sqrt{\Delta}= \frac {|k|\pi}2$ one can obtain
\be
C(\k)= \frac {\k\pi}2\, \frac {{\rm cosh}(\frac {\k\pi}2)}
{{\rm sinh}(\frac {\k\pi}2)}\label{C(k)}
\ee
where the + sign has been chosen in taking the square root. This is due to
the fact that also $C$ is a positive matrix (its main diagonal is given
by the sequence $2,4,6,...$, while the off--diagonal terms have alternating
signs and decreasing absolute value). Anyhow, the eigenvalue
of $C$ could be calculated directly with the same method used for
$\tilde A(\k)$. But since this type of calculations will be used repeatedly
in the ghost sector, we will dispense with doing it in this case.

\subsection{The wedge states}

The matter wedge states are squeezed states (like in the RHS of
eq.(\ref{matterwedge})) that satisfy the recursion relation
$|n+1\rangle = |n\rangle \star |2\rangle$, for $n\geq 1$,
where $|2\rangle$ is identified
with the $|0\rangle$ vacuum. It is easy to prove that, as a consequence,
the corresponding matrices $S_n$ have to satisfy the recursion
relations, \cite{Furu,Kishimoto},
\be
T_{n+1}= X\frac {1-T_n}{1-T_nX},\label{recur}
\ee
where $T_n= \hat C S_n$ and $X=\hat C V^{rr}$,
$V^{rr}$ being matrices of Neumann
coefficients of the three strings vertex. 
Moreover their normalization factors ${\cal N}_n$ have to satisfy
\be
{\cal N}_n\, {\cal K} \, \det \left(1-T_n X\right)^{-\frac 12} =
{\cal N}_{n+1}\label{normrecur}
\ee
where ${\cal K}$ is some constant to be determined.
gv
We have now all at hand to show that these recursion relations are indeed satisfied
by the squeezed states constructed from the LHS of eq.(\ref{matterwedge})
via the formula (\ref{KPmatter}), and therefore prove the equivalence
(\ref{matterwedge}).
In order to make the comparison with (\ref{recur},\ref{normrecur}),
we multiply everything in (\ref{commcase})
by $\hat C$, the twist matrix, and evaluate $\tilde \alpha = \hat C \alpha$
at $t=t_n\equiv -\frac{n-2}2$. We expect to find (from now on, in this section,
we actually denote with the symbols of matrices their eigenvalues)
\be
{T_n}\equiv \hat C S_n = -2\,\tilde \alpha\left(-\frac{n-2}2\right)\label{iden}
\ee
i.e.
\be
{T_n}=2 \tilde A
\frac {{\rm sinh}\left(\sqrt{\Delta}\,\frac{n-2}2\right)}
{ \sqrt{\Delta}\, {\rm cosh} \left(\sqrt{\Delta}\,\frac{n-2}2\right)
+C\, {\rm sinh}\left(\sqrt{\Delta}\,\frac{n-2}2\right)}\label{Tn}
\ee
In particular, we should find
\be
T_3 = X,\quad \quad  \lim_{n\to \infty} T_n =
\frac {2\tilde A}{C+\sqrt{\Delta}}\equiv T\label{identif}
\ee
where $S=\hat C T$ is the sliver matrix.
Replacing the continuous basis and using
\be
\tilde A=- \frac {\k\pi}{4\, {\rm sinh}\left(\frac {\k\pi}2\right)},\quad\quad
C= \frac {\k\pi}2 \, {\rm coth}\left(\frac {\k\pi}2\right)\label{A&C}
\ee
this is precisely what one gets.

More generally one must prove the recursion relation (\ref{recur})
or its solution
\be
T_n= \frac {T+(-T)^{n-1}}{1-(-T)^n}\label{recursolution}
\ee
To avoid cumbersome calculations let us proceed as follows.
We use the condensed notations
\be
{\rm sinh}\left(\sqrt{\Delta}\,\frac{n-2}2\right)= {\rm sh}_n,
\quad\quad {\rm cosh}\left(\sqrt{\Delta}\,\frac{n-2}2\right)= {\rm ch}_n
\0
\ee
and the trigonometric formulas
\be
&&{\rm sh}_{n+1} = {\rm sh}_n {\rm ch}_3+{\rm ch}_n {\rm sh}_3\0\\
&&{\rm ch}_{n+1} = {\rm ch}_n {\rm ch}_3+{\rm sh}_n {\rm sh}_3\0
\ee
Then we have in particular
\be
&&T_1= - 2\tilde A \frac {{\rm sh}_3}{\sqrt{\Delta} \,{\rm ch}_3-C\,{\rm sh}_3}
\label{T1}\\
&&T_2=0\0\\
&&T_3 = 2\tilde A \frac {{\rm sh}_3}{\sqrt{\Delta} \,{\rm ch}_3+C\,{\rm sh}_3}
\label{T3}
\ee
Replacing the trigonometric formulas inside $T_{n+1}$
we get
\be
T_{n+1} = 2\tilde A \frac {{\rm sh}_n{\rm ch}_3+{\rm ch}_n {\rm sh}_3}
{\sqrt{\Delta} ({\rm ch}_n {\rm ch}_3+{\rm sh}_n {\rm sh}_3)+
C({\rm sh}_n {\rm ch}_3+{\rm ch}_n {\rm sh}_3)}\label{Tn+1}
\ee
We want to compare this with the RHS of eq.(\ref{recur})
\be
T_3\frac {1-T_n}{1-T_nT_3}= 2 \tilde A
\frac {\frac {C-2\tilde A}{\sqrt{\Delta}}{\rm sh}_n{\rm sh}_3+{\rm ch}_n
{\rm sh}_3}
{\sqrt{\Delta} ({\rm ch}_n {\rm ch}_3+{\rm sh}_n {\rm sh}_3)+
C({\rm sh}_n {\rm ch}_3+{\rm ch}_n {\rm sh}_3)}\label{RHS}
\ee
Therefore (\ref{Tn+1}) and (\ref{RHS}) coincide if
\be
\frac {C-2\tilde A}{\sqrt{\Delta}}{\rm sh}_3= {\rm ch}_3\label{T1=id}
\ee
Notice that this is exactly the condition for which $T_1=1$, as it should be.
Replacing the eigenvalues of the previous subsection it is elementary
to prove (\ref{T1=id}).

As for the normalization constants ${\cal N}_n$ they must satisfy
(\ref{normrecur}). Set
\be
\eta_n = 2 \int_0^{t_n}\,dt\,{\rm tr} (\alpha B)
=2\int_0^{t_n}dt\,{\rm tr}(\tilde \alpha \tilde A)\label{etan}
\ee
and identify
\be
{\cal N}_n = e^{\eta_n} \0\label{etanNn}
\ee
Now, inserting the results of the previous subsection one can verify
that (\ref{normrecur}) is satisfied, and determine ${\cal K}$. The result
is expressed in terms of untraced quantities $\hat \eta_n$ such that
$\eta_n = {\rm tr} (\hat \eta_n)$. Setting $y=\frac {\pi \k}2$,
\be
\hat\eta_n= -\frac {n-2}4\,y\, \coth (y) + \frac 12 \log \left( \frac {\sinh(y)}
{\sinh \left(y \frac n2\right)}\right)\label{hatetan}
\ee
and ${\cal K}= e^{\eta_3}$. The latter result is a consequence of
imposing ${\cal N}_2=1$, which is required if we want the wedge state
$|2\rangle$ to coincide with the vacuum $|0\rangle$.

Taking the traces corresponds to integrating over
$\k$ from $-\infty$ to $\infty$ and introducing an infinite factor.
These objects need to be regularized. But this
is a well--known problem when normalizing squeezed states.

\section{The ghost sector. Propaedeutics}

We would now like to repeat the above analysis also for the ghost sector,
and prove that\footnote{The RHS of (\ref{ghostwedge}) is of course meant to
represent the ghost wedge states. It should be stressed
that defining star product of these states is a rather non--trivial
problem in the oscillator formalism. We
postpone the treatment of this issue to a forthcoming paper, \cite{BMST} and,
for the time being, 
the reader is advised to think that everything works as in the matter sector.}
\be
|n\rangle = e^{ - \frac {n-2}2\left({\cal L}^{(g)}_0 +
{\cal L}_0^{(g)\dagger}\right)}|0\rangle
= {\cal N}_n\, e^{ c^\dagger S_n b^\dagger}|0\rangle \label{ghostwedge}
\ee
(to avoid a proliferation of symbols we use the same ones as in the matter
case, but it should be clear that from now on we deal only with ghosts; so $S_n$
is the wedge matrix for the ghost sector, etc.).

A crucial question for the ghost sector is what normal ordering we use.
In string theory
the conventional normal ordering is the one implied by the vacuum
$c_1|0\rangle$, where $|0\rangle$ is the SL(2,R) invariant vacuum,
identified by
the regularity of the ghost fields $c(z)$ and $b(z)$ at $z=0$. We call
{\it conventional} this normal ordering, while we call {\it natural} the
normal ordering implied by the choice of the $|0\rangle$ vacuum.
More explicitly, in  the natural n.o. $c_{1},c_0,c_{-1},c_{-2},...$ are
creation operators and the complementary ones annihilation operators,
while in the conventional n.o. $c_0, c_{-1},c_{-2},...$ are creation
operators and the complementary one annihilation operators (with
the conjugate prescriptions for the $b_n$'s).

We will see in the forthcoming sections that the natural normal ordering
fits in a natural way in the task of solving the KP equations for the
ghost sector, while there may be some problems with the conventional
normal ordering.

This section is devoted to collect some preparatory material for the
ghost sector.

\subsection{Virasoro generators with the conventional n.o.}

The Virasoro generators for the ghost sector (for some basic information
concerning ghosts in string field theory
see \cite{Samu,GJ1,GJ2,leclair,aref,tope}), normal ordered in the
conventional way, are
\be
L_n^{(g)} &=& \sum_{k\geq 1} (2n+k)\, b_k^\dagger c_{k+n} -\sum_{k\geq 0}
(n-k) \,c_k^\dagger b_{n+k} -\sum _{k=1}^{n}(n+k)\,c_k b_{n-k}\label{Lng}\\
L_0^{(g)} &=& \sum_{k\geq 1}k\, c_k^\dagger b_k +
\sum_{k\geq 1}k\,b_k^\dagger c_k-1
\label{L0g}\\
L_{-n}^{(g)} &=& \sum_{k\geq 1} (k-n)\, b_{n+k}^\dagger c_{k} +
\sum_{k\geq 0}
(2n+k) \,c_{k+n}^\dagger b_{k}+
\sum _{k=0}^{n-1}(n+k)\,c_k^\dagger b_{n-k}^\dagger
\label{L-ngh}
\ee
Therefore we can write
\be
{\cal L}_0^{(g)}+{\cal L}_0^{(g)\dagger} &=& 2L_0^{(g)} + \sum_{n=1}^\infty
\frac {2(-1)^{n+1}}
{4n^2-1} (L_{2n}^{(g)}+ L_{-2n}^{(g)})\0\\
&=& c^\dagger \,{ A}\,b^\dagger +
c^\dagger {C}\, b + b^\dagger { D} c - c\, {B}\,b-2\label{L0gh'}
\ee
where
\be
{A}_{pq} &=& \sum_{n=1}^\infty \ell_n (n+p) \delta_{p+q,n} =(2p+q)\,
\ell_{p+q},\quad\quad p\geq 0,\quad q\geq 1\label{Ag}\\
{B}_{pq} &=&  \sum_{n=1}^\infty \ell_n (n+p) \delta_{p+q,n} =(2p+q)\,
\ell_{p+q},\quad\quad p\geq 1,\quad q\geq 0 \label{Bg}\\
{C}_{pq} &=& \sum_{n=1}^\infty \ell_n [(p-n) \delta_{p+n,q} +
(2n+q) \delta_{q+n,p}] + 2\, p\,\delta_{p,q}\0\\
&=& (2p-q)\,\ell_{|q-p|},\quad\quad p\geq 0,\quad q\geq 0 \label{Cg}\\
{D}_{pq} &=& \sum_{n=1}^\infty \ell_n [(q-n) \delta_{q+n,p} +
(2n+p) \delta_{p+n,q}] + 2\, p\,\delta_{p,q}\0\\
&=& (2q-p)\,\ell_{|q-p|},\quad\quad p\geq 1,\quad q\geq 1\label{Dg}
\ee
In view of these formulas it is natural to introduce two types of indices:
 the unbarred ones, $p, q$, running from $0$ up to $+\infty$ and the
barred ones $\bar p, \bar q$, running from $1$ up to $+\infty$.
The above matrices will carry either type of indices:
$C=\{C_{pq}\}$ is a square `long--legged' matrix, $D=\{D_{\bar p\bar q}\}$
is square `short--legged'. The matrix $A= \{A_{p\bar q}\}$
($B=\{B_{\bar p,q}\}$) is instead left--long--legged and right--short--legged
(right--long--legged and left--short--legged). We will refer to this kind
of matrices as {\it lame} matrices\footnote{If the matrices in question were
finite the appropriate term would
be rectangular matrices, but semi--infinite matrices can always be seen as
square, provided we relabel the indices. It is clear that in this context what
matters is the bases these matrices are applied to. For this reason we use
the {\it ad hoc} term `lame'.}.

We notice that in the overlapping span of the indices we have
$A=B$ and $C=D^T$. Actually we will learn in due course that $A$ and $B$
($C$ and $D^T$) represent the same operator applied to two different bases.

\subsection{The Virasoro generators with the natural n.o.}

The Virasoro generators for the ghost sector in the natural normal
ordering are ($n>0$)
\be
L_n^{(g)} &=& \sum_{k\geq 2} (2n+k)\, b_k^\dagger c_{k+n} -\sum_{k\geq -1}
(n-k) \,c_k^\dagger b_{n+k} -\sum _{k=2}^{n+1}(n+k)\,c_k b_{n-k}
\label{Lngnat}\\
L_0^{(g)} &=& \sum_{k\geq -1}k\, c_k^\dagger b_k +
\sum_{k\geq 2}k\,b_k^\dagger c_k
\label{L0gnat}\\
L_{-n}^{(g)} &=& \sum_{k\geq 2} (k-n)\, b_{n+k}^\dagger c_{k} +
\sum_{k\geq -1}
(2n+k) \,c_{k+n}^\dagger b_{k}+
\sum _{k=-1}^{n-2}(n+k)\,c_k^\dagger b_{n-k}^\dagger
\label{L-nghnat}
\ee
Therefore we can write
\be
{\cal L}_0^{(g)}+{\cal L}_0^{(g)\dagger} &=& 2L_0^{(g)} + \sum_{n=1}^\infty
\frac {2(-1)^{n+1}}
{4n^2-1} (L_{2n}^{(g)}+ L_{-2n}^{(g)})\0\\
&=& c^\dagger \,{A}\,b^\dagger +
c^\dagger {C}\, b + b^\dagger {D} c - c\, {B}\,b\label{L0ghnat}
\ee
where
\be
{A}_{p\bar q} &=& (2p+\bar q)\ell_{|p+\bar q|},\label{Agnat}\\
{B}_{\bar p q} &=&(2\bar p+q)\ell_{|\bar p+q|}, \label{Bgnat}\\
{C}_{pq} &=& (2p-q)\,\ell_{|q-p|}\label{Cgnat}\\
{D}_{\bar p \bar q} &=& (2\q-\p)\,\ell_{|\q-\p|}\label{Dgnat}
\ee
where again we have two types of indices, but, at variance with the previous
section, now the unbarred ones run from $-1$
up to $+\infty$, while the barred ones run from $2$ to $+\infty$. Again
we have different type of matrices. $C$ is a long--legged square matrix,
$D$ is a short--legged square matrix. $A$ and $B$ are `lame' matrices.
We also remark that again for overlapping indices
we have ${A}={B}$ and ${C}={D}^T$. At times we will refer to $A$ and $B$
as $A$--type matrices and to $C$ and $D^T$ as $C$--type matrices.

\subsection{KP equations for the ghosts}

Our purpose is to prove a relation similar to (\ref{KPmatter})
\be
e^{t(c^\dagger A\, b^\dagger + c^\dagger C\, b + b^\dagger D c - c\, B\,b)}
=e^\eta \, e^{c^\dagger \alpha\, b^\dagger} \, e^{c^\dagger \gamma\, b}\,
e^{b^\dagger \delta\, c} \, e^{c\, \beta\,b}\label{expo}
\ee
where $\alpha,\beta,\gamma,\delta,\eta$ depend on $t$ and they are matrices
of the same type as $A,B,C,D$ (that is, $\alpha$ is right--long--legged and
left--short--legged matrix, etc.). Proceeding as in KP, we differentiate
both sides, commute to the left and obtain the equations
\be
&& A= \dot {\alpha} - C\, \alpha - \alpha\, D^T -
\alpha\,B\alpha\label{Ag1}\\
&& C = -\alpha\, B +\frac {d e^{\gamma}}{dt} e^{-{\gamma}}\label{Cg1}\\
&&D= -\alpha^T B^T+ \frac {d e^{\delta}}{dt} e^{-\delta}\label{Dg1}\\
&&B = e^{-\delta^T } \dot {\beta}\, e^{-\gamma}\label{Bg1}\\
&&0= \dot \eta +
{\rm Tr}\left(B\,\alpha\right)\label{etag1}
\ee
In this paper we are interested only in eq.(\ref{Ag1}) and (\ref{etag1}).
Notice the ordering of factors in the trace in (\ref{etag1}).

\subsection{Some very useful identities}

Before we proceed we need to derive a few basic identities that will turn up
in all the subsequent calculations. In the following we use the symbol
$\CA$ and $\CC$ to represent square matrices that are related to $A$ and $C$
but do not coincide exactly with them (they correspond to $A$ and $C$ in the
range $1,2,\ldots, \infty $), but they give rise
to sort of universal relations, which can be used in the rest of the paper
with just few modifications.

Let first $p,q$ be both odd, then one can prove the following
\be
(\CA \CC)_{pq} &=& \sum_{k=1}^\infty (2p+k)(2k-q)\,
\ell_{p+k}\,\ell_{|k-q|}\0\\
&=&
4 (-1)^{\frac {p+q}2}\sum_{l=0}^\infty
\frac {(2p+2l+1)(4l-q+2)}{((p+2l+1)^2-1)((2l-q+1)^2-1)}
\0\\
&=&2 (-1)^{\frac {p+q}2} \left[-\left(\frac{(q-2)(2p+q-1)}{q(p+q) (p+q-2)}
+ \frac{(p-1)(2p+q+2)}{p(p+q) (p+q-2)}\right)\right.\0\\
&&+ \left.\frac{4p^2+3pq+q^2-4}{(p+q)((p+q)^2-4)}\left(\psi( \frac p2)-
\psi(- \frac q2)\right)\right]\label{ACpq}
\ee
and
\be
(\CC \CA)_{pq} &=& \sum_{k=1}^\infty (2p-k)(2k+q)\,
\ell_{q+k}\,\ell_{|k-p|}\0\\
&=&
4 (-1)^{\frac {p+q}2} \sum_{l=0}^\infty
\frac {(2p-2l-1)(4l+q+2)}{((p-2l-1)^2-1)((2l+q+1)^2-1)}\0\\
&=& 2(-1)^{\frac {p+q}2} \left[-\left(\frac{(q+2)(2p+q+1)}{q(p+q) (p+q-2)} +
\frac{(p+1)(2p+q-2)}{p(p+q) (p+q-2)}\right)\right.\0\\
&&+ \left.\frac{4p^2+3pq+q^2-4 }{(p+q)((p+q)^2-4)}\left(\psi( - \frac p2)-
\psi( \frac q2)\right)\right]\label{CApq}
\ee
Using $\psi(  \frac p2)-\psi( - \frac p2)= -\frac 2p$ we get exactly
\be
(\CA\CC)_{pq}=(\CC\CA)_{pq} \label{pqodd}
\ee
This is valid only for $p,q>1$, since for $p=q=1$ there is a singularity
in the previous formulas. This case must be calculated separately.
One gets
\be
[\CA,\CC]_{11} = \sum_{l=0}^\infty \frac {16\,l(l+1)}{(2l+1)(2l+3)(4l^2-1)} =
\frac {\pi^2}8\label{AC11}
\ee

Let now $p,q$ be both even. We have
\be
(\CA \CC)_{pq} &=& \sum_{l=1}^\infty (2p+2l)(4l-q)\, \ell_{p+2l}\,\ell_{|2l-q|}\0\\
&=&
8 (-1)^{\frac {p+q}2} \sum_{l=1}^\infty
\frac {(p+l)(4l-q)}{((p+2l)^2-1)((2l-q)^2-1)}\0\\
&=&2 (-1)^{\frac {p+q}2}
\left[-\left(\frac{(q-2)(2p+q-1)}{(q-1)(p+q) (p+q-2)} +
\frac{(p-1)(2p+q+2)}{(p+1)(p+q) (p+q-2)}\right)\right.\0\\
&&+ \left.\frac{(2p-q)^2-4}{(p+q)((p+q)^2-4)}\left(\psi(\frac 12 + \frac p2)-
\psi(\frac 12 - \frac q2)\right)\right]\label{ACpqeven}
\ee
and
\be
(\CC \CA)_{pq} &=& \sum_{l=1}^\infty (2p-2l)(4l+q)\, \ell_{q+2l}\,\ell_{|2l-p|}\0\\
&=&
8 (-1)^{\frac {p+q}2} \sum_{l=1}^\infty
\frac {(p-l)(4l+q)}{((p-2l)^2-1)((2l+q)^2-1)}\0\\
&=&2 (-1)^{\frac {p+q}2}
\left[-\left(\frac{(q+2)(2p+q-2)}{(q+1)(p+q) (p+q-2)} +
\frac{(p+1)(2p+q-2)}{(q-1)(p+q) (p+q-2)}\right)\right.\0\\
&&+ \left.\frac{(2p-q)^2-4}{(p+q)((p+q)^2-4)}\left(\psi(\frac 12 - \frac p2)-
\psi(\frac 12 + \frac q2)\right)\right]\label{CApqeven}
\ee
Using $\psi(\frac 12 -\frac p2)=\psi(\frac 12 +\frac p2)$ for $p$ even,
we get
\be
(\CA \CC-\CC \CA)_{pq}=8 (-1)^{\frac {p+q}2}\,\frac {pq}{(p^2-1)(q^2-1)}=2\,
u_p\,u_q\label{AC-CApq}
\ee
where $u_p = p\,\ell_p$.

Another important relations is the one concerning $\CC^2- \CA^2$.
Proceeding as in Appendix A one gets, for odd $p,q$,
\be
(\CC^2-\CA^2)_{pq}= \frac{\pi^2}2 \left( (p^2-2)\,\delta_{p,q}+ \frac 12 p(p-1)
\,\delta_{q,p+2}
+\frac 12 p(p+1) \,\delta_{p,q+2}\right)\label{C2-A2odd}
\ee
while for even $p,q$
\be
(\CC^2-\CA^2)_{pq}&=& 8 \frac{pq\, (-1)^{\frac{p+q}2}}{(p^2-1)(q^2-1)}
\label{C2-A2even}\\
&&+\frac{\pi^2}2 \left( (p^2-2)\delta_{p,q}+ \frac 12 p(p-1) \,\delta_{q,p+2}
+\frac 12 p(p+1) \,\delta_{p,q+2}\right)\0
\ee
All these equations will be repeatedly applied in the sequel.

\section{Integrating the KP equations in the ghost sector. Natural n.o.}

In this section our purpose is to integrate the KP equations in the ghost
sector. We concentrate on the case of the natural n.o., which provides
a clean framework to solve the problem and postpone till the end of the paper
the discussion of the problems with conventional n.o.
In order to be able to tackle the integration of the KP equation
we need some remarkable identities satisfied by the
$A,B,C,D$ matrices, analogous to the ones in the matter sector.

\subsection{Commuting matrices}

First let us show that $AD^T= CA$.
Suppose $p$ and $\bar q$ are even. Then
\be
(A D^T)_{p \bar q} &=& \sum_{\bar k=2}^\infty A_{p\bar k}\,
D^T_{\bar k \bar q}=
\sum_{l=2}^\infty (2p+2l)(4l-\bar q) \ell_{|\bar q+2 l|} \,
\ell_{|2l-p|}\0\\
&=& 8 (-1)^{\frac {p+\bar q}2} \sum_{l=1}^\infty \frac {(p+l)(4l -\bar q)}
{((p+2l)^2-1)((2l-\bar q)^2-1)}\label{ADTpqeenat}
\ee
having set $\k= 2l$. We notice that the RHS of this equation coincide with
(\ref{ACpqeven}). Similarly
\be
&&(C A)_{p \bar q} = \sum_{\bar k=-1}^\infty C_{p k}\,
A_{k \bar q}=
\sum_{l=0}^\infty (2p-2l)(4l+\bar q) \ell_{|p+2 l|} \,
\ell_{|2l-\bar q|}\0\\
&=& 8 (-1)^{\frac {p+\bar q}2}\left[ \sum_{l=1}^\infty \frac {(p-l)(4l +\bar q)}
{((p-2l)^2-1)((2l+\bar q)^2-1)} + \frac {p\bar q}{(p^2-1)(\bar q^2-1)}\right]
\label{CApqeenat}
\ee
The first term in the square bracket on the RHS of this equation is nothing but
eq.(\ref{CApqeven}).
Putting everything together and using eq.(\ref{AC-CApq}), we find
\be
(A D^T - C A)_{p\bar q} = 8 (-1)^{\frac {p+\bar q}2}
\left[\frac {p\bar q}{(p^2-1)(\bar q^2-1)}-
\frac {p\bar q}{(p^2-1)(\bar q^2-1)} \right]=0\label{ADT-CAnat}
\ee
This derivation becomes singular in the case $p=0,\bar q=2$. This case has to
be handled separately. One easily gets
\be
(A D^T)_{02}= -\frac 14, \quad\quad (C A)_{02} =-\frac 14\label{02casenat}
\ee

Similarly for $p$ and $\bar q$ odd we have
\be
(A D^T)_{p \bar q} &=& \sum_{\bar k=3}^\infty A_{p\bar k}\,
D^T_{\bar k \bar q}=4 (-1)^{\frac {p+\bar q}2}
\left[\sum_{l=0}^\infty \frac {(2p+2l+1)
(4l -\bar q +2)}
{((p+2l+1)^2-1)((2l-\bar q+1)^2-1)}\right.\0\\
&-&\left. \frac {(2p+1)(2-\bar q)}{((p+1)^2-1)
((q-1)^2-1)} \right]
\label{ADTpqoonat}
\ee
and
\be
(C A)_{p \bar q} &=& \sum_{\bar k=-1}^\infty C_{p k}\,
A_{k \bar q}= 4(-1)^{\frac {p+\bar q}2}
\left[ \sum_{l=0}^\infty \frac {(2p-2l-1)(4l +\bar q+2)}
{((p-2l-1)^2-1)((2l+\bar q+1)^2-1)}\right.\0\\
 &+&\left. \frac {(2p+1)(\bar q-2)}{((p+1)^2-1)
((q-1)^2-1)} \right]\label{CApqoonat}
\ee
having set $k = 2l+1$. The first term in square brackets in (\ref{ADTpqoonat})
coincides with eq.(\ref{ACpq}), while the corresponding term in
the RHS of (\ref{CApqoonat}) coincides with (\ref{CApq}). Therefore they are
equal. It follows that the RHS of (\ref{ADTpqoonat}) equals the RHS of
(\ref{CApqoonat}). The previous analysis does not hold for $p=-1$ and $q=3$,
but one can prove directly that
\be
(A D^T)_{-1,3}=-\frac 14 =(C A)_{-1,3} \label{-13case}
\ee
Therefore we can conclude that
\be
A D^T = C A \label{ADT=CAnat}
\ee
which is the appropriate `commutation relation' for these kind of matrices.
In the same way one can prove that
\be
BC = D^T B\label{BC=DTBnat}
\ee

It follows at once from these results that, for instance,
\be
C A B = A D^T B = A B C \label{CABnat}
\ee

\subsection{Integration of the KP equations}

We are now ready to integrate the KP equations.
We concentrate on
\be
\dot {\alpha}=A +C {\alpha} +{\alpha} D^T +{\alpha}  B
{\alpha}\label{KPnat}
\ee
with the initial condition $\alpha(0)=0$.
The solution to
\be
\dot \eta = -{\rm Tr}\left(B\,\alpha\right)\label{etanat}
\ee
is obvious once we know $\alpha(t)$. Let us assume that
\be
\alpha \,D^T= C \,\alpha\label{commhypnat}
\ee
and multiply (\ref{KPnat}) from the right by $B$. Defining
$ \theta = \alpha B $, eq.(\ref{KPnat}) becomes
\be
\dot \theta = AB + 2 C\theta +\theta^2\label{newKPnat1}
\ee
since $C \theta = \theta C$ on the basis of (\ref{commhypnat}).
Since $C AB=ABC$, eq.(\ref{newKPnat1}) can be now easily
integrated like a numerical equation,
provided also $\theta AB=AB\theta$. The result is
\be
\theta(t) = A B \frac {{\rm sinh}\left(\sqrt{C^2-AB}\,t\right)}
{ \sqrt{C^2-AB}\, {\rm cosh} \left(\sqrt{C^2-AB}\,t\right)
-C\, {\rm sinh}\left(\sqrt{C^2-AB}\,t\right)}
\label{thetat}
\ee
It is easy to see that the commutativity hypotheses made above are all
satisfied. Notice that $\theta(t)$ is a square long--legged matrix.
From $\theta$
we can extract $\alpha$ by multiplying from the right by the inverse of $B$
(concerning the invertibility of $B$, see the end of the next section).
Let us denote the resulting solution by $\alpha_1$:
\be
\alpha_1(t) = \frac {{\rm sinh}\left(\sqrt{C^2-AB}\,t\right)}
{ \sqrt{C^2-AB}\, {\rm cosh} \left(\sqrt{C^2-AB}\,t\right)
-C\, {\rm sinh}\left(\sqrt{C^2-AB}\,t\right)}\,A\label{1stalphat}
\ee
This solution is left--long--legged and right--short--legged.

If we multiply (\ref{KPnat}) from the left by $B$, we can get another
solution,
say $\lambda(t) = B\,\alpha(t)$. Proceeding in the same way as above
we find
\be
\lambda(t) = BA \frac {{\rm sinh}\left(\sqrt{(D^T)^2-BA}\,t\right)}
{ \sqrt{(D^T)^2-BA}\, {\rm cosh} \left(\sqrt{(D^T)^2-BA}\,t\right)
-D^T\, {\rm sinh}\left(\sqrt{(D^T)^2-BA}\,t\right)}
\label{lambdat}
\ee
This solution is represented by a square
short-legged matrix. So we can extract another form of the solution to the
KP equation
\be
\alpha_2(t) = A \frac {{\rm sinh}\left(\sqrt{(D^T)^2-BA}\,t\right)}
{ \sqrt{(D^T)^2-BA}\, {\rm cosh} \left(\sqrt{(D^T)^2-BA}\,t\right)
-D^T\, {\rm sinh}\left(\sqrt{(D^T)^2-BA}\,t\right)}
\label{2ndalphat}
\ee
However commuting $A$ to the right from (\ref{2ndalphat}) we get (\ref{1stalphat})
and viceversa. Therefore the two solutions are one and the same.
This unique solution will be referred to as $\alpha(t)$.

\subsection{Other useful formulas}

A crucial role is clearly played in the above formulas by $(D^T)^2-BA$ and
$C^2-AB$. They turn out to be matrices of a very simple form.
Indeed using (\ref{C2-A2odd},\ref{C2-A2even}) we can prove that,
for odd $p,q$
\be
(C^2-AB)_{pq}&=& \sum_{k=-1} C_{pk} C_{kp}-\sum_{k=2} A_{pk} B_{kp}
=(\CC^2-\CA^2)_{pq}+ C_{p0} C_{0q}\0\\
 &=&\frac{\pi^2}2 \left( (p^2-2)\,\delta_{p,q}+ \frac 12 p(p-1)
\,\delta_{q,p+2}
+\frac 12 p(p+1) \,\delta_{p,q+2}\right)\label{C2-A2oddnat}
\ee
and for even $p,q$
\be
(C^2-AB)_{pq}&=& \sum_{k=-1} C_{pk} C_{kp}-\sum_{k=2} A_{pk} B_{kp}=
(\CC^2-\CA^2)_{pq} + C_{p,-1} C_{-1,q}+A_{p,1}A_{1,q}\0\\
&=&\frac{\pi^2}2 \left( (p^2-2)\delta_{p,q}+ \frac 12 p(p-1) \,\delta_{q,p+2}
+\frac 12 p(p+1) \,\delta_{p,q+2}\right)\label{C2-A2evennat}\
\ee
i.e., in general,
\be
(C^2-AB)_{pq}=\frac{\pi^2}2 \left( (p^2-2)\delta_{p,q}+ \frac 12 p(p-1)
\,\delta_{q,p+2}
+\frac 12 p(p+1) \,\delta_{p,q+2}\right)\label{C2-ABnat}
\ee

These equalities hold for square long--legged matrices. As above
one can prove in a similar way
\be
((D^T)^2-BA)_{\bar p \bar q}= \frac{\pi^2}2 \left( (\bar p^2-2)
\delta_{\bar p,\bar q}+ \frac 12 \bar p(\bar p-1) \,\delta_{\bar q,\bar p +2}
+\frac 12 \bar p (\bar p+1) \,\delta_{\bar p,\bar q+2}\right)
\label{D2-ABnat}
\ee
for square short--legged matrices.

\section{Diagonalization. Natural n.o.}

We have seen in the previous section that the $A$--type and $C$--type matrices
commute (although not in the form square matrices do, but in a more elaborate
form fit for rectangular--type matrices). This has allowed us to integrate
the KP equations. Now we would like to proceed further in the direction of
proving the equivalence (\ref{ghostwedge}). It proves to be very difficult
if not impossible to do so in the discrete matrix language used so far.
As in the matter case it is convenient to pass to a continuous basis.
The $A$ and $C$--type matrices are however not easy to diagonalize, it is
convenient to choose a simpler operator that commutes with them
and has a nondegenerate spectrum. This is $K_1=L_1+L_{-1}$, as
suggested by Rastelli, Sen and Zwiebach, \cite{RSZ1}.

In the ghost sector with the natural normal ordering we get
\be
K_1= \sum_{p,q\geq -1} c_p^\dagger\, G_{pq} \,b_q + \sum_{p,q\geq 2}
b_{\bar p}^\dagger\,H_{\bar p\bar q}\,
c_{\bar q}- 3c_2\,b_{-1}\label{K1gh}
\ee
where
\be
&&G_{pq}= (p-1) \delta_{p+1,q} + (p+1)\delta_{p-1,q},\0\\
&&H_{\bar p\bar q}= (\bar p+2) \delta_{\bar p+1,\bar q}
+ (\bar p-2)\delta_{\bar p-1,\bar q}\label{GF}
\ee
Therefore $G$ is a square long--legged matrix and $H$ a square
short--legged one. In the common overlap
we have $G=H^T$.

Since we want to transfer the relevant information contained in $K_1$, in
its oscillator form, to infinite matrices and replace the effect of oscillator
anticommutation relations with matrix operation, it is clear that the last
term in the RHS of (\ref{K1gh}) is an awful nuisance. Fortunately
the effect of this term can be incorporated in the matrices $G$ and $H$
provided we suitably enlarge them. More precisely we can write
\be
K_1= \sum_{\dot p q} c_{\dot p}^\dagger\, G_{\dot p q} \,b_q +
\sum_{\tilde p,\q}
b_{\tilde p}^\dagger\,H_{\tilde p\bar q}\,
c_{\bar q} \label{K1gh'}
\ee
where $G,H$ are now lame matrices
\be
&&G_{\dot p q}= (\dot p-1) \delta_{\dot p+1,q} + (\dot p+1)\delta_{\dot
p-1,q},\0\\
&&H_{\tilde p \q}= (\tilde p+2) \delta_{\tilde p+1,\q}
+ (\tilde p-2)\delta_{\tilde p-1,\bar q}\label{GF'}
\ee
where $\dot p$ runs from -2 to $\infty$ and $\tilde q$ from 1 to $\infty$.
We will refer to these indices as the {\it stretched} ones and to the
corresponding matrix enlargement as {\it stretching}.
Since this implies two additional terms corresponding to $G_{-2,-1}$
and $H_{1,2}$, it would seem that we are double--counting, but this is not
the case if we use the following prescription:
in each of this new terms in $K_1$ we have one daggered and one
undaggered operator, for instance $b_{-1}$ appears in
the $H_{1,2}$ term as the daggered operator $b_1^\dagger$ and has nontrivial
anticommutaion relation with the undaggered operator $c_1$
(not with $c_{-1}^\dagger$).
That is the anticommutation rules are nontrivial only between daggered and
conjugate undaggered operators.
This allows us to transfer all the content of the oscillator anticommutation
relations to the product of matrices. In order to see that this is the correct
prescription one can check that these rules reproduce the correct commutation
relations of $K_1$ with the conformal $b(z)$ and $c(z)$ fields (see below).

In general in conformal field theory, for a primary field of weight $h$, we have
\be
[{\cal L}_0+{\cal L}_0^\dagger, \phi(z)] = ({\rm arctg}(z) +{\rm arctg}(1/z))
\left((1+z^2)\,\partial \phi(z)+2\,z\,h\,\phi(z)\right)\label{primary}
\ee
With these formulas one can compute the commutator between $K_1$ and
${\cal L}_0+{\cal L}_0^\dagger$. Using the property
$\frac{\partial}{\partial z}({\rm arctg}(z) +{\rm arctg}(1/z))=0$, it is
elementary to see that
\be
[[{\cal L}_0+{\cal L}_0^\dagger,K_1]\,,\,\phi(z)]=0\0
\ee
for any primary field. This must be true in particular for $c(z)$ and $b(z)$.

Therefore, returning to the oscillator representation of
${\cal L}_0+{\cal L}_0^\dagger$
in terms of matrices and of $K_1$ in terms of $G$ and $H$ (\ref{K1gh}),
we expect to find
\be
0=[K_1,{\cal L}_0+{\cal L}_0^\dagger]=\quad  c^\dagger
(GA + A H^T)b^\dagger -c(B G+ H^TB)b+
c^\dagger [G,C]b+ b^\dagger[H,D]c \0
\ee
We indeed find
\be
(GA+A H^T)_{p\q}= (2p+\q)(p+\q+2)\ell_{p+\q+1}+
(2p+\q)(p+\q-2)\ell_{p+\q-1}=0\0
\ee
We remark that in this equation the internal summations are the ordinary ones,
over a `long' index $n$ in $GA$ and a `short' index $\bar n$ in $A H^T$.
Similarly
\be
[G,C]_{pq}=(2p-q)(p-q-2) \ell_{|p-q-1|}+(2p-q)(p-q+2)\ell_{|p-q+1|}=0\0
\ee
The internal dummy index is an ordinary `long' one in $GC$, while it is
a stretched one $\dot n$ in $CG$. This is due to the previous stretching
of $G$ and consequent stretching of $C$. This carries into the game a new
term (which is precisely the one that is produced by the third
term in the RHS of (\ref{K1gh})).

In a similar way we can prove $(BG+H^TB)_{\bar p \bar q} =0$
and $[H,D]_{\bar p \bar q}=0$. This is true only because of the
extension of the internal summations due to the
stretching of $H, G$ and corresponding stretching of $B,D$,
following the above recipe.

We notice that
since the twist matrix $\hat C$ anticommutes with $G$ and $H$
we get the `commutation' rules (fit for lame matrices)
\be
(G\tilde A -\tilde A H^T)_{p\q}=0=(\tilde B G- H^T\tilde B)_{\p \q}
\label{lamecomm}
\ee

The stretching does not affect the matrix $A$, while $B$ is both left and
right--stretched, $C$ and $D$ are right--stretched.
We would like to notice that matrix stretching introduced above
carries a stretching also in the commutators and consequently
new rows or columns. For instance we have
$(G\tilde A -\tilde A H^T)_{-2,\q}\neq 0$ or $(GA+A H^T)_{p 1}\neq 0$,
and similar relations for the other commutators. This nonzero terms
however do not interfere with our diagonalization because they are outside
the range of the bases to which we will apply these commutation relations.

It is important to remark that this technical complication is not intrinsic
to the problem we want to solve, i.e. diagonalizing the solution of
the KP equations, but rather to the technique we use to solve it. In fact
our true aim is to prove eq.(\ref{eigenforA}) below and the like. Since
we cannot do it directly, because of the complexity of the problem, we
resort to the operator $K_1$, which can be easily diagonalized and has
a nondegenerate spectrum, and use the fact that operators that commute
with it must have the same eigenvectors. It is in taking this detour that
we meet the complication of having to use stretched matrices. Fortunately,
as we will see, this implies only some awkwardnesses in the formulas,
but disappears in the final results.

\subsection{The ${\cal H}^{(2)}$ space}

We call ${\cal H}^{(2)}$ the space spanned by the eigenvectors of $K_1$
belonging
to the weight 2 basis, \cite{Belov1,Belov2}. These are obtained as follows.
Let us start from the action of  $K_1$ on the weight 2 field $b(z)$
\be
{\cal K}_1 b(z) &\equiv&[K_1,b(z)] = (1+z^2)\partial b(z) +4 z\,b(z)
= - \sum_n\left((n+1)\,b_{n-1} +(n-1) b_{n+1}\right) z^{-n-2}\0
\ee
On the other hand, on the operator side,
\be
[K_1,b(z)] &=&\left[ c^\dagger G\,b+b^\dagger H\, c\, ,\,
\sum_{n\geq -1} b_n\, z^{-n-2}
+ \sum_{n>1} b_n^\dagger z^{n-2}\right]\0\\
&&= - \sum_n \left((n-1) \,b_{n+1} +(n+1)b_{n-1}\right) \, z^{-n-2}
\label{K1bop}
\ee
We stress that in order for this and the corresponding equation for
the $c$ field to be true we have to use stretched $H,G$ matrices.

Therefore ${\cal K}_1$ and $K_1$ with the adjoint action, generate the
same conformal transformation
on the $z$--plane. Now let us integrate the differential equation
\be
{\cal K}_1 f^{(2)}_\k(z)  = (1+z^2)\partial f^{(2)}_\k(z) +4 z\,f_\k(z)
= \kappa f^{(2)}_\k(z)\0
\ee
The result is the generating function
\be
f^{(2)}_\k(z) = \left(\frac 1{1+z^2}\right)^2 \, e^{\k \arctan (z)}=
1+\k z +\left(\frac {\k^2}2 -2\right) z^2+\ldots\label{genf}
\ee
The unnormalized basis (weight 2 basis) is given by
\be
f^{(2)}_\k(z) = \sum_{n=2} V_n^{(2)}(\k) \, z^{n-2}\label{bbasis}
\ee
i.e.
\be
V_n^{(2)}(\k) =\frac 1{2\pi i}\oint dz\,
\frac {e^{\k \arctan(z)}}{(1+z^2)^2} \frac 1{z^{n-1}}
\label{intrepofbbasis}
\ee
Notice the different labeling with respect to \cite{Belov1,Belov2}.
Following \cite{Belov1,Belov2}, (see also Appendix B), we normalize
the eigenfunctions as follows
\be
\tilde V_n^{(2)}(\k)= {\sqrt{A_2(\k)}} V_n^{(2)}(\k)
\label{nbbasis}
\ee
where
\be
  A_2(\k) = \frac {\k (\k^2+4)}
{\sinh \left(\frac {\pi \k}2\right)}\0
\ee

\subsection{The ${\cal H}^{(-1)}$ space}

We call ${\cal H}^{(-1)}$ the space spanned by the eigenvectors of $K_1$
belonging to the weight -1 basis. It is obtained in the same way as above
by replacing the $b(z)$ with the $c(z)$ field. In short, we start with
\be
{\cal K}_1 f^{(-1)}_\k(z)  = (1+z^2)\partial f^{(-1)}_\k(z) -2 z\,f_\k(z)
= \kappa f^{(-1)}_\k(z)\0
\ee
which is dictated by the transformation of the $c$ field,
and integrate the differential equation. The result is
the generating function
\be
f^{(-1)}_\k(z) = (1+z^2) \, e^{\k \arctan (z)}=
1+\k z +\left(\frac {\k^2}2 +1\right) z^2+\ldots\label{f-1}
\ee
The unnormalized basis is given by
\be
f_\k^{(-1)}(z) = \sum_{n=-1} V_n^{(-1)}(\k) \, z^{n+1}\label{cbasis}
\ee
i.e.
\be
V_n^{(-1)}(\k) =\frac 1{2\pi i}\oint dz\,(1+z^2)\,
{e^{\k \arctan(z)}}\frac 1{z^{n+2}}
\label{V-1n}
\ee
The normalized one is
\be
\tilde V_n^{(-1)}(\k)=\sqrt{A_{-1}(\k)} V_n^{(-1)}(\k),
\quad\quad \sqrt{A_{-1}(\k)} = \frac 12 {\cal P} \frac 1{\k}
\frac {\sqrt{A_2(\k)}}{\k^2+4}\label{normV-1n}
\ee
where ${\cal P}$ denotes the principal value.

Before proceeding let us briefly comment about the two bases $V_n^{(2)}(\k)$
and $V_n^{(-1)}(\k)$ (a more complete account is presented in Appendix B,
based on the results of \cite{Belov1,Belov2,BeLove}). We can
promote $V_n^{(2)}(\k)$ to a
complete orthonormal basis in ${\cal H}^{(2)}$ by multiplying them by
suitable normalization
constants $N_n^{(2)}$. The so obtained new states, $\hat V_n^{(2)}$, are
however not anymore eigenstates of $H^T$ (see below). We can do the same
with $V_n^{(-1)}(\k)$ and get a complete orthonormal basis
$\hat V_n^{(-1)}(\k)$, but the latter are not eigenfunctions of
$G$. This may seem to be a drawback, but in fact what we need is another
property: the weight 2 and -1 bases, $\tilde V_n^{(2)}(\k)$ and
$\tilde V_n^{(-1)}(\k)$ are biorthogonal, see Appendix B.

\subsection{Diagonalization of $K_1$}

It is now easy to verify that, as expected, the matrices representing
$K_1$ are diagonal in the above bases. Let us start with the weight 2 basis.
\be
\sum_{n=2}^\infty H^T_{pn}\, V_n^{(2)}(\k) &=&
\frac 1{2\pi i} \oint
dz\, \frac {e^{\k \arctan(z)}}{(1+z^2)^2}\, \frac 1{z^{n-1}}
\left( (p-1)\,\delta_{p+1,n}
+(p+1)\,\delta_{p-1,n}\right)\0\\
&=& \frac 1{2\pi i} \oint
dz\,  \frac {e^{\k \arctan(z)}}{(1+z^2)^2} \left(\frac {p-1}{z^p}
+\frac {p+1}{z^{p-2}}\right)\0\\
&=& -\frac 1{2\pi i} \oint dz\,  e^{\k \arctan(z)} \frac d{dz} \left(
\frac 1{1+z^2} \frac 1{z^{p-1}}\right)\0\\
&=& \k\, V^{(2)}_p(\k)\label{GVp}
\ee
as expected. In the weight -1 basis we have
\be
\sum_{n=-1}^\infty  V_n^{(-1)}(\k)\,G_{nq} &=& \frac 1{2\pi i}
\oint dz (1+z^2) \,e^{\k \arctan (z)}\0\\
&&\cdot\sum_{n=-1}^\infty \frac 1{z^{n+2}} \left( (n-1) \delta_{n+1,q}
+(n+1)\delta_{n-1,q}\right)\0\\
&=& -\frac 1{2\pi i}
\oint \,e^{\k \arctan (z)}\frac d{dz} \left((1+z^2)^2\frac 1{z^{q+2}}\right)
\0\\
&=& \k \,  V_q^{(-1)}(\k)\label{V-1nG}
\ee
In other words $V^{(2)}_n(\k)$ right--diagonalizes $H^T$, while
$V^{(-1)}_n(\k)$ left--diagonalizes $G$. We remark that the stretching
of $G$ and $H$ does not affect the above equations.

\subsection{Diagonalization of $C^2-AB$}

The weight 2 basis right--diagonalizes $(D^T)^2-BA$. Using (\ref{D2-ABnat})
we get
\be
&&\sum_{n=2}^\infty ((D^T)^2-BA)_{pn}\, V_n^{(2)}(\k) =\0\\
&=&\frac {\pi^2}2 \frac 1{2\pi i} \oint dz\,
\frac {e^{\k \arctan(z)}}{(1+z^2)^2}
\left( \frac {p^2-2}{z^{p-1}}+ \frac 12 \frac {p^2-p}{z^{p+1}}
+\frac 12 \frac {p^2+p}{z^{p-3}}\right)\0\\
&=&\frac {\pi^2}2\, \frac 1{2\pi i}  \oint dz\,
\frac {e^{\k \arctan(z)}}{(1+z^2)^2}\cdot
\frac 12\, (1+z^2)^2\, \frac d{dz} \left( (1+z^2)\,\frac d{dz}
\left(\frac 1{1+z^2} \,
\frac 1{z^{p-1}}\right)\right)\0\\
&=&\frac {\pi^2\k^2}4\,\frac 1{2\pi i} \oint dz\,
\frac {e^{\k \arctan(z)}}{(1+z^2)^2} \frac 1{z^{p-1}}=
 \frac {\pi^2\k^2}4\, V_p^{(2)}(\k)\label{D2BAVp}
\ee
Similarly the weight -1 basis left--diagonalizes $C^2-AB$. Using
(\ref{C2-ABnat}) one gets
\be
\sum_{n=-1}^\infty  V_n^{(-1)}(\k)\,(C^2-AB)_{nq} &=&
\frac {\pi^2}4\frac 1{2\pi i}
\oint dz (1+z^2) \,e^{\k \arctan (z)}\0\\
&&~~~\cdot\left(2\frac {q^2-2}{z^{q+2}}+
\frac {(q-2)(q-3)} {z^q} +\frac {(q+2)(q+3)}{z^{q+4}} \right) \0\\
&=&\frac {\pi^2}4\frac 1{2\pi i}
\oint dz (1+z^2) \,e^{\k \arctan (z)}\0\\
&&~~~\cdot \frac 1{1+z^2} \frac d{dz} \left(
(1+z^2) \,\frac d{dz} \left((1+z^2)^2 \frac 1{z^{q+2}}\right)\right)\0\\
&=& \frac {\pi^2}4 \, \k^2  \,V_p^{(-1)}(\k)\label{C2-ABcbase}
\ee

\subsection{Eigenvalue of $\tilde A$ in the weight 2 basis}

It is of course incorrect to speak about the eigenvalue of a lame matrix
like $\tilde A$, because $\tilde A$ transform short--legged vectors into
long--legged ones. Therefore the eigenvalue equation, strictly speaking,
cannot be satisfied. What we really mean here is the following. We
wish to prove that
\be
&&\sum_{\q=2}^\infty \tilde A_{\p \q} \, V_{\q}^{(2)}(\k) = {\mathfrak a}(\k)
V_{\p}^{(2)}(\k), \label{eigenforA}\\
&&\sum_{\q=2}^\infty \tilde A_{a \q}\, V_{\q}^{(2)}(\k) =0,\quad\quad
a=-1,0,1\label{eigenA=0}
\ee
This is precisely what we need in order to demonstrate the result in the next
section. Eq.(\ref{eigenA=0}) is proven in Appendix D1. Next
let us call $\EA$ the square submatrix
$\EA_{\p \q}=A_{\p \q}=B_{\p \q}$. Thanks to (\ref{eigenA=0}) it is
easy to realized that, when applied to the basis $V_{\p}^{(2)}(\k)$
the relation (\ref{lamecomm}),
$G\tilde A -\tilde A H^T=0$,  reduces to $[\tilde \EA, H^T]=0$. This and
(\ref{GVp}) are enough
to conclude that $\tilde \EA$ is diagonal in the weight
2 basis. This is precisely eq.(\ref{eigenforA}). From this we also see
that the stretching introduced at the beginning of this section is
completely harmless.

Now let us compute the eigenvalue ${\mathfrak a}(\k)$,
following the method of \cite{RSZ1}. We must have
\be
\sum_{n=2}^\infty\,\tilde \EA_{2,n}\,V_n^{(2)}(\k)=
{\mathfrak a}(\k)\,V_2^{(2)}(\k)={\mathfrak a}(\k)
\label{eigeneq}
\ee
because $V_2^{(2)}(\k)=1$. Now, for $n=2l$ even,
we have
\be
\tilde \EA_{2,n}\equiv
\tilde A_{2,n}=(-1)^{\frac n2} \left(\frac 3{n+1}- \frac
1{n+3}\right)\label{A2n}
\ee
The strategy consists in: ({\it step
1}) writing a differential equation for ${\mathfrak a}(\k)$ and integrating
it; then, ({\it step 2}), expressing the integral in terms of
hypergeometric functions. Finally, ({\it step 3}), using the
properties of the latter to simplify the result. We will work out
this case in some detail as a sample of several computation of the same kind.

\noindent {\bf Step 1}

Let us write
\be
F(z)&=& \sum_{l=1}^\infty \frac{(-1)^l }{2l+1} V_{2l}^{(2)} (\k)
z^{2l+1}\label{F(z)}\\
G(z)&=& \sum_{l=1}^\infty \frac{(-1)^l }{2l+3} V_{2l}^{(2)} (\k)
z^{2l+3}\label{G(z)}
\ee
Then
\be
\frac {dF}{dz}&=&
\sum_{l=1}^\infty (-1)^l  V_{2l}^{(2)} (\k)\, z^{2l}=
-\frac{z^2}2\left(f_\k^{(2)}(iz)+f_\k^{(2)}(-iz)\right)\label{dF(z)}\\
\frac {dG}{dz}&=& \sum_{l=1}^\infty (-1)^l  V_{2l}^{(2)} (\k)\,
z^{2l+2}=
-\frac{z^4}2\left(f_\k^{(2)}(iz)+f_\k^{(2)}(-iz)\right)\label{dG(z)}
\ee
where $f_\k^{(2)}(z)$ was defined above. Let us set
\be
H(z)=3F(z)-G(z)\label{H(z)}
\ee
We get
\be
\frac{dH}{dz} =-\frac {3z^2-z^4}{2}\,
\left(f_\k^{(2)}(iz)+f_\k^{(2)}(-iz)\right)\0
\ee
Therefore
\be
{\mathfrak a}(\k)\equiv H(1) = - \int_0^1 dz \, h(z)
\cosh \left( \k \tg^{-1}(iz)\right)\label{H(1)}
\ee
where
\be
h(z) = \frac {3 z^2-z^4}{(1-z^2)^2} \label{h(z)}
\ee
We will need
\be
\tg^{-1}(iz) = i \,\th^{-1}(z)= \frac i2\,
\log \left( \frac {1+z}{1-z}\right)\0
\ee
As it is evident from this equation in the integrand of (\ref{H(1)})
there is a branch point at 1,
therefore this expression of the integral is formal and we have to specify
its meaning.

\noindent {\bf Step 2}

We do it in the following way. We rewrite
\be
H(1) &=& \frac 12 \int_0^1 dz \left(
2z^2\,(1+z)^{\zeta-2} (1-z)^{-\zeta-2} + z^2\,(1+z)^{\zeta-1} (1-z)^{-\zeta-1}
\right)+\0\\
&&+(\k\rightarrow -\k) = \Theta(\k) + \Theta(-\k)\label{Theta}
\ee
where $\zeta= \frac {i\k}{2}$.

Using the basic integral representation of the hypergeometric function
(\ref{hyper1})
\be
\Theta(\k) =-\frac 1{\zeta(1-\zeta)(1+\zeta)}\left[ 2\,
F\left(2-\zeta,3;2-\zeta;-1\right)-
\frac {1+\zeta}{2-\zeta} \, F\left(1-\zeta,3;3-\zeta;-1\right)\right]\label{Thetak}
\ee
Here $F(a,b;c;z)\equiv{}_2\! F_1(a,b;c;z)$. For other properties of
the hypergeometric functions, see Appendix C.

\noindent {\bf Step 3}

The strategy now consists in bringing the hypergeometric functions to
the form
(\ref{hyper2}), (\ref{hyper3}) or (\ref{hyper0}). To this end one uses
(\ref{hyper4}) or (\ref{hyper5}). One then uses, for instance, the
properties of the $\EG$ special function
\be
\EG(1+z) +\EG(z) = \frac 2z,\quad\quad \EG(1-z)+\EG(z)=
\frac {2\pi}{\sin(\pi z)}\0
\ee
Using (\ref{hyper0}) one gets $F(2-\zeta,3;2-\zeta;-1)
= \frac 18$.
The other expression is more complicated. Using three times (\ref{hyper5})
on $F\left(1-\zeta,3;3-\zeta;-1\right)$ one gets a sum of terms of the form
(\ref{hyper0}) or (\ref{hyper3}). Finally
\be
\Theta(\k)= -\frac 1{4(\zeta-\zeta^3)} +\frac 12+ \frac 1{4\zeta} +
\frac {\zeta}2 \, \EG(1-\zeta)\label{thetak2}
\ee
Then
\be
{\mathfrak a}(\k)\equiv H(1)&=&\Theta(\k)+ \Theta(-\k)=
1+\frac {\zeta}2 \left(\EG(1-\zeta)-\EG(1+\zeta)\right)\0\\
&=& 1+ \frac {\zeta}2 \left(\EG(1-\zeta)+\EG(\zeta)-\frac 2{\zeta}\right)=
\frac {\pi \k}2 \frac 1{\sinh \left(\frac {\pi \k}2\right)}\label{H1fin}
\ee

We have done an independent check of this result by considering the
equation
\be
\sum_{n=2}^\infty\,\tilde \EA_{3,n}\,V_n^{(2)}(\k)=
{\mathfrak a}(\k)\,V_3^{(2)}(\k)=\k\,{\mathfrak a}(\k)
\label{eigeneq2}
\ee
Analytically this case is very complicated, but numerically one can easily show
that it gives the same result (\ref{H1fin}).

\subsection{The eigenvalues of $\tilde B, D^T$ in the weight 2 basis,
and $\tilde A$ in the weight -1 basis}

In a very similar way one can prove that $\tilde A$, when applied from the right
to the weight -1 basis has the same eigenvalue (\ref{H1fin}). On the other
hand the eigenvalue of $D^T$ in the
weight 2 basis is given by ${\mathfrak c}(\k)$
\be
{\mathfrak c}(\k) = \frac{\pi \k}2 \, \coth \left(\frac{\pi \k}2\right)
\label{gamma}
\ee
All these results are derived in Appendix D. As for $\tilde B$, due to
eq.(\ref{eigenA=0}), it is easy to show that the product
$\tilde B \tilde A$ boils down to $\EA^2$ when acting on the weight 2 basis.
Therefore the eigenvalue of $\tilde B$ (in the sense explained at the
beginning of the previous section) is ${\mathfrak a}(\k)$. It follows
that, if we consider for instance the solution (\ref{lambdat}) applied
to the weight 2 basis, it makes sense to multiply it
from the left by $\tilde B^{-1}$ and end up with (\ref{2ndalphat}).

\section{Ghost wedge states}

We are now in the position to draw the conclusions of our long analysis on the
ghost sector. Let us consider the solution (\ref{2ndalphat}) to the
KP equations. We apply it to the weight 2 basis from the left.
We remark that due to the form of  (\ref{2ndalphat}), since $A$ is not
stretched\footnote{In solving the KP equations we must use stretched
matrices, but it is easy to realize that the solution is unstretched.}
the stretching of $B$ and $D$ disappears, and we can safely use
(\ref{D2BAVp}): $(D^T)^2-BA$ is diagonal with eigenvalue
$\frac {\pi^2 \k^2}4$. Moreover $D^T$ is diagonal in this basis
with eigenvalue ${\mathfrak c}(\k)$, and $\tilde A$ too (in the sense
explained above) with eigenvalue
\be
{\mathfrak a}(\k)\,=\, \frac {\pi\k}{2\,
\sinh\left(\frac{\pi \k}2\right)}
\0
\ee
Notice the different sign and factor of 2, with respect to the matter sector.
So $\alpha_2(t)=\alpha(t)$ is diagonal too in the weight 2 basis
with eigenvalue given by
(\ref{2ndalphat}) where all the matrices have been replaced by the
corresponding eigenvalues. Let us call $\alpha(\k,t)$ the
resulting eigenvalue.

Let us return to eq.(\ref{ghostwedge}). In this equation we have to compute
in particular $c^\dagger \alpha b^\dagger$. Using the formalism of Appendix
B, we can write this term as follows
\be
c^\dagger \alpha(t) b^\dagger &=&
\sum_{n=-1, m=2} c^\dagger_n \alpha_{nm}(t) b_m^\dagger
=  \sum_{n=-1, m=2}^\infty \int d\k\, d\k'\,\tilde c(\k)
\tilde V_n^{(-1)}(\k)
\tilde \alpha_{nm}(t)  \tilde V_m^{(2)}(\k')\, b(\k')\0\\
&=& \sum_{n=2}^\infty \int d\k\, d\k'\, \tilde c(\k)\tilde V_n^{(-1)}(\k)
\tilde \alpha(\k,t) \tilde V_n^{(2)}(\k')b(\k')=
 \int d\k\,\tilde c(\k)\tilde A(\k)\, b(\k)\label{cABdagger1}
\ee
(see Appendix B for the definitions of $\tilde c(\k)$ and $b(\k)$, and the
biorthogonality relation used in (\ref{cABdagger1})).
The same can be done for any power of $\alpha(t), \tilde A$ and $D^T$ and
$\Delta= (D^T)^2-BA$. Therefore, from now on we will simply deal with the
corresponding eigenvalues.

Now let us apply this to the ghost wedge states. The ghost wedge
states are squeezed states  such as those in the RHS of
eq.(\ref{ghostwedge}), that satisfy the recursion relation \be
|n+1\rangle = |n \rangle \star |2\rangle, \quad\quad n\geq
1\label{ghostrecur} \ee where $|2\rangle$ is identified with the
ghost vacuum $|0\rangle$. This recursion relation in turn implies
the recursion relation \be T_{n+1}= X\frac
{1-T_n}{1-T_nX},\label{recurgh} \ee for the matrices $T_n=\hat C
S_n$, and \be {\cal N}_n\, {\cal K} \, \det \left(1-T_n X\right) =
{\cal N}_{n+1} \label{normrecurgh} \ee for the normalization
factors ${\cal N}_n$, with ${\cal N}_2=1$. In order to prove these
relations for squeezed states one has to introduce the three
strings vertex for the ghost part, ($X$ in (\ref{recurgh}) is
$\hat C V^{11}$, the latter being a matrix of Neumann coefficients
of the three strings vertex). All this will appear in a
forthcoming paper, \cite{BMST}
\footnote{By considering the weight-0 primary $Y(z)=\frac12 \partial^2c\partial c c(z)$, the recursion relation for normalizations,
 (\ref{normrecurgh}) can be
understood as the result of the overlap between the
wedge states $| n\rangle$ and the ghost number 3 states $|\hat m\rangle= e^{\frac {2-m}2 \left({\cal L}_0+{\cal L}_0^\dagger\right)}Y(0)|0\rangle$,
in the form $\langle\hat2|n+1\rangle=\langle\hat3|n\rangle$. The origin of this relation is given by
$\langle\hat 2|n+1\rangle=\langle\hat 2| \left(|n\rangle*|2\rangle\right)$ and it does not depend on the actual location of $Y(z)$ at the boundary.
}.
{\it For the time being
we suppose we can replace the matrices in
(\ref{recurgh},\ref{normrecurgh}) by their eigenvalues in the
weight 2 basis. This will also be justified in \cite{BMST}}.

Under this hypothesis let us set out to prove that the states
$|n\rangle$ given by (\ref{ghostwedge}) satisfy (\ref{recur}) and
(\ref{normrecur}).
This sentence has to be unambiguously understood. First we notice that
$T_2=0$, which is consistent with $|2\rangle$ being identified with the
vacumm $|0\rangle$ and ${\cal N}_2=1$. Now proving  (\ref{recur})
means proving two things
\be
T_3=X \label{firsteq}
\ee
and
\be
T_{n+1}= T_3\frac {1-T_n}{1-T_n T_3},\label{secondeq}
\ee

As we did in the matter case, in the rest of this section we use the
matrix symbols to actually represent the corresponding eigenvalues
in the weight 2 basis. We should have
\be
{T_n}\equiv \hat C S_n = \tilde \alpha\left(-\frac{n-2}2\right)\label{idengh}
\ee
where, as usual, a tilde represents a twisted matrix
$\tilde \alpha=\hat C \alpha$. In detail
\be
{T_n}= -\tilde A
\frac {{\rm sinh}\left(\sqrt{\Delta}\,\frac{n-2}2\right)}
{ \sqrt{\Delta}\, {\rm cosh} \left(\sqrt{\Delta}\,\frac{n-2}2\right)
+D^T\, {\rm sinh}\left(\sqrt{\Delta}\,\frac{n-2}2\right)}\label{Tngh}
\ee
where $\sqrt{\Delta} =\frac {\pi |\k|}2$, while
\be
\tilde A= \frac {\k\pi}{2\, {\rm sinh}\left(\frac {\k\pi}2\right)},\quad\quad
D^T= \frac {\k\pi}2 \, {\rm coth}\left(\frac {\k\pi}2\right)\label{A&Cgh}
\ee

Comparing with the expression of the wedge states in the matter sector,
we see that the expression of $T_n$ in terms of $\k$ are exactly
the same, because the relative minus sign and factor of 2 in the definition
of $T_n$ are compensated by the change of sign and the factor of 2 in the
expression of the eigenvalue of $\tilde A$. Therefore, without repeating
the demonstration of section 2.5, we know that (\ref{secondeq}) is true.

We recall that (\ref{secondeq}) can be solved by
\be
T_n= \frac {T+(-T)^{n-1}}{1-(-T)^n}\label{recursolutiongh}
\ee
for some matrix $T$. This matrix is easily identified to be
$T\equiv T_\infty$ (this makes sense because the absolute value (of the
eigenvalue) of $T$ turns out to be $<1$:
$T= -e^{-\frac {|\kappa|\pi}2}$). But, from the defining
eq.(\ref{ghostrecur}), $T$ must represent the sliver.
Therefore it is related to $X$ by
\be
X= \frac T{T^2-T+1}\label{inversesliver}
\ee
This proves eq.(\ref{firsteq}). For completeness it remains to show that
the matrices $S_n= \hat C T_n$ are the same that represent the wedge surface 
states as squeezed states, see for instance \cite{Schnabl05}.
{\it This will be done in \cite{BMST}}.

As in the matter sector the normalization constants ${\cal N}_n$ must also
satisfy a recursion relation
where ${\cal K}$ is a constant to be determined by
the condition ${\cal N}_2=1$ (i.e. ${\cal K}={\cal N}_3$).
We have
\be
\eta_n = - \int_0^{t_n}\,dt\,{\rm tr} (B\alpha)\label{etangh}
\ee
Using the simple relations of these quantities with their matter counterpart
and identifying
\be
{\cal N}_n = e^{\eta_n} \0\label{etanNngh}
\ee
we can verify that (\ref{normrecurgh}) is satisfied with ${\cal N}_2=1$
and ${\cal K}= {\cal N}_3$.
We remark that in order to define the trace in (\ref{etangh}) the
completeness relation (\ref{bi-completeness}) is needed.

In this section we have solved the problem by applying $\alpha(t)$ to the
weight 2 basis from the left. We believe that the same result can be obtained
by applying it to the weight -1 basis from the right, even though the
calculations turn out to be more involved. For this reason
we have started the relevant analysis (see, for instance, App. D3), but
we have not completed it.

\section{Twisted ghost sector}

In this section, for completeness we deal with the twisted ghost sector,
\cite{GRSZ1}.
This sector, due to its features very close to the those of the matter sector,
allows for a simplified treatment.
The twist changes the geometrical nature of the ghost fields
so that $b$ gets weight 1 and $c$ weight 0.

The unordered Virasoro generators for the twisted ghosts are
\be
L_n= - \sum_k \, k b_{n+k} c_{-k}\0
\ee
Once they are normal ordered
\be
L_n^{(g)} &=& \sum_{k\geq 1} (n+k)\, b_k^\dagger c_{k+n} +\sum_{k\geq 0}
k \,c_k^\dagger b_{n+k} -\sum _{k=1}^{n}k\,\,c_k b_{n-k}\label{twLng}\\
L_0^{(g)} &=& \sum_{k\geq 1}k\, c_k^\dagger b_k +
\sum_{k\geq 1}k\,b_k^\dagger c_k
\label{twL0g}\\
L_{-n}^{(g)} &=& \sum_{k\geq 1} k\, b_{n+k}^\dagger c_{k} +
\sum_{k\geq 0}
(n+k) \,c_{k+n}^\dagger b_{k}+\sum _{k=0}^{n-1}k\,c_k^\dagger b_{n-k}^\dagger
\label{L-ng}
\ee
Therefore, for the ghost part, we can write
\be
{\cal L}_0+{\cal L}_0^\dagger &=& 2L_0^{(g)} + \sum_{n=1}^\infty
\frac {2(-1)^{n+1}}
{4n^2-1} (L_{2n}^{(g)}+ L_{-2n}^{(g)})\0\\
&=& c^\dagger \,{\bf A}\,b^\dagger +
c^\dagger {\bf C}\, b + b^\dagger {\bf D} c - c\, {\bf B}\,b\label{L0gh}
\ee
where we have introduced boldface matrices for a reason that
will be clear in a moment,
\be
{\bf A}_{pq} &=& \sum_{n=0}^\infty \ell_n p \delta_{p+q,n} =p\,
\ell_{p+q},\quad\quad p\geq 0,q\geq 1\label{twAg}\\
{\bf B}_{pq} &=&  \sum_{n=0}^\infty \ell_n p \delta_{p+q,n} =-p\,
\ell_{p+q},\quad\quad p\geq 1,q\geq 0\label{twBg}\\
{\bf C}_{pq} &=& \sum_{n=0}^\infty \ell_n [p \delta_{p+n,q} +
(n+q) \delta_{q+n,p}] + 2\, p\,\delta_{p,q}\0\\
&=& p(\ell_{q-p}+\ell_{p-q})= p\ell_{|p-q|}, \quad\quad p\geq 0,q\geq 0
\label{twCg}\\
{\bf D}_{pq} &=& \sum_{n=0}^\infty \ell_n [q \delta_{q+n,p} +
q\delta_{p+n,q}] + 2\, p\,\delta_{p,q}\0\\
&=& q(\ell_{q-p}+\ell_{p-q})=q\ell_{|p-q|},\quad\quad p\geq 1,q\geq 1
\label{twDg}
\ee
The matrices $\A,\B,\C,\D$ have the following structure.
\be
&&\A = \left(\begin{matrix} 0 &\mid& 0\\
                         ---&\mid&---\\
                         0&|& A\\
\end{matrix}\right),\quad\quad\quad
\C= \left(\begin{matrix} 0 &\mid&  0\\
---&\mid&---\\
                         \u&|& C\\
 \end{matrix}\right) \label{A,C}\\
&&\B = \left(\begin{matrix} 0 &\mid& 0\\
                         ---&\mid&---\\
                         \u&|& A\\
\end{matrix}\right),\quad\quad\quad
\D= \left(\begin{matrix} 0 &\mid&  0\\
---&\mid&---\\
                         0&|& C^T\\
 \end{matrix}\right) \label{B,D}
\ee
where $\u=\{u_p\}$ and $u_p=p\,\ell_p$. The matrices $\A,\B,\C,\D$ are
referred to as the large matrices, while $A$ and $C$ will be called small
matrices.

In analogy with what was done in section 4, we can immediately check that
$\A\D^T-\C\A=0$ $\Longleftrightarrow$ $[A,C]=0$, while
\be
\B\,\C-\D^T\,\B =
\left(\begin{matrix} 0 &\mid& 0\\
                         --------&\mid&-------------\\
                         A\u&|& AC-CA\\
\end{matrix}\right)\label{AC-CAtwgh}
\ee
On the other hand there is a direct relation with the corresponding matter
matrices, compare with \cite{MM},
\be
A^{(m)} = \frac 12 E^{-1} { A}\,E,\quad\quad
C^{(m)} =E^{-1} { C}\,E,\label{relACMatTwgh}
\ee
where $E_{nm} = \sqrt{n} \delta_{nm}$ and the label $^{(m)}$ of course denote
the matter counterpart. It follows in particular that
\be
[A,C]=0\label{[A,C]}
\ee

Using this relation with the matter part we also get
\be
(C^2-A^2)_{pq} &=& \left(E( C^{(m) 2}-4 A^{(m)2})E^{-1}\right)_{pq}\0\\
&=&\frac {\pi^2}2 \left( p^2\delta_{p,q} +\frac 14 p(p+q)
(\delta_{p,q+2}+\delta_{q,p+2})\right)\label{C2A2tw}
\ee
This relation does not hold if we include the zero mode (we do not have
the analogs of the relations in section 4.3). However, in the untwisted ghost
sector we do not have to repeat the derivation. Here life is easier.

\subsection{The KP equation}

Now we would like to integrate the KP equation
\be
\dot{\hat\alpha}= \A + \C \hat \alpha + \hat\alpha \D^T+ \hat\alpha\,
\B\hat\alpha
\label{twghalpha}
\ee
where $\hat \alpha$ represent the large matrix solution, while the small one
will be denoted by $\alpha$. Looking at the relation (\ref{expo}) we can
guess the form to be expected for $\hat\alpha$. It can only be
\be
\hat\alpha = \left(\begin{matrix} 0 &\mid& 0\\
                         ---&\mid&---\\
                         \f&|& \alpha\\
\end{matrix}\right),\label{hatalpha}
\ee
which breaks down to the equations
\be
&&  \dot \f =  (C+\alpha A)\f\label{alphaeq2}\\
&&\dot{\alpha}= A +\{\alpha,C\}+ \alpha\, A\,\alpha\label{alphaeq3}
\ee
It is consistent to set $\f=0$ (this is actually the solution one gets by
solving (\ref{twghalpha}) recursively starting from $\A$).
Therefore only the second equation remains and can be solved precisely as
in the matter case since the small matrices $A$ and $C$ commute:
\be
\alpha(t) = A \frac {{\rm sinh}\left(\sqrt{C^2-A^2}\,t\right)}
{ \sqrt{C^2-A^2}\, {\rm cosh} \left(\sqrt{C^2-A^2}\,t\right)
-C\, {\rm sinh}\left(\sqrt{C^2-A^2}\,t\right)}
\label{twghalphasol}
\ee

\subsection{Representation of $K_1$}

Let us consider the $K_1=L_1+L_{-1}$ operator in the twisted ghost case.
We get
\be
K_1= \sum_{p,q\geq 0} c_p^\dagger\, G_{pq} \,b_q +
\sum_{p,q\geq 1}b_p^\dagger\,H_{pq}\,
c_q+c_1\,b_0\label{K1twgh}
\ee
where
\be
&&G_{pq}= p (\delta_{p+1,q} + \delta_{p-1,q}),\0\\
&&H_{pq}= q( \delta_{p+1,q} + \delta_{p-1,q})=G_{qp}\label{twGF}
\ee
So $G$ is a long--legged square matrix, while $H$ is a short--legged square
one. However this distinction is pointless here because the
nonvanishing  0--th column present in $G$ is immaterial in what follows.
Also in this, like in the untwisted case, we have the problem
of the additional term in the RHS of eq.(\ref{K1twgh}) and
we can remedy in the same way, by absorbing it in the first two
terms at the price of introducing stretched matrices. $G$ becomes
a left--stretched matrix with the first index running from -1 to $\infty$,
and $H$ a left--stretched matrix with the first index starting from
0 instead of 1, and retaining nontrivial anticommutation rules only
between daggered and undaggered operators.

All the conditions for $K_1$ to commute with ${\cal L}_0+{\cal L}_0^\dagger$
are satisfied. In fact it is easy to prove that
\be
[G,\C]_{pq}= p(p-q+2)\,\ell_{|p-q+1|}+p(p-q-2)\,\ell_{|p-q-1|}=0,
\quad\quad p,q\geq 0\0
\ee
provided in $\C G$ the summation index runs from $-1$ to $\infty$.
Similarly one can prove that $G\A +\A H^T=0$ and $[H,\D]=0$, which
do not require any stretching. Finally $\B G+H^T\B=0$, provided
in $\B G$ the summation index is stretched. Once again we will realize
that stretching does not have practical consequences except
to help us in the process of diagonalization.

\subsection{The weight 1 basis}

For weight 1 ghost field $b$ we have
\be
{\cal K}_1 \, b(z) =[K_1,b(z)]= (1+z^2)\partial b(z) + 2zb(z)\0
\ee
Integrating the equation
\be
{\cal K}_1\, f_\k^{(1)}(z)= \k\, f_\k^{(1)}(z)\0
\ee
we get
\be
f_\k^{(1)}(z)\equiv \sum_{n=1}^\infty \,V_n^{(1)}(\k) z^{n-1} =
\frac 1{1+z^2} e^{\k\, {\arctan} (z)}=1+\ldots\label{fk1}
\ee
So, in particular, $V_1^{(1)}=1$, and
\be
V_n^{(1)}(\k)= \frac 1{2\pi i} \oint dz\, \frac 1{1+z^2} \,
\frac {e^{\k\,{\arctan}(z)}}{z^n}\label{V1noint}
\ee
As a verification, for instance, we have
\be
\sum_{n=1}H^T_{pn}\, V_n^{(1)}(\k) &=&\frac 1{2\pi i} \oint dz\, \frac 1{1+z^2} \,
e^{\k{\,\arctan}(z)} p\, \sum_{n=1}^\infty (\delta_{p+1,n}+\delta_{p-1,n})\,
\frac 1{z^n}\0\\
&=&- \frac 1{2\pi i} \oint dz\,
e^{\k{\,\arctan}(z)}\frac d{dz} \left(\frac 1{z^p}\right)\0\\
&=&  \k\, V_p^{(1)}(\k) \label{HTV1n}
\ee
and, similarly,
\be
\sum_{n=1} \, V_n^{(1)}(\k) H_{nq}= \k\, V_q^{(1)}(\k)\0
\ee

In the same way we can compute
\be
\sum_{n=1}^\infty (C^2-A^2)_{pn} \, V_n^{(1)}(\k) &=&
\frac 1{2\pi i} \oint dz\, \frac 1{1+z^2} \,
e^{\k{\,\arctan}(z)}\0\\
&&\cdot \frac {\pi^2}2 \sum_{n=1}^\infty\left( p^2\delta_{p,n} +\frac 14 p(p+n)
(\delta_{p,n+2}+\delta_{n,p+2})\frac 1{z^n}\right)\0\\
&=&\frac {\pi^2}4\,\frac 1{2\pi i} \oint dz\, \frac
{e^{\k{\,\arctan}(z)}}{1+z^2}\,(1+z^2) \frac d{dz}\left( (1+z^2)\frac d{dz}
\frac 1{z^p}\right)\0\\
&=& \frac {\pi^2\k^2}4\,\frac 1{2\pi i} \oint dz\, \frac
{e^{\k{\,\arctan}(z)}}{1+z^2} \,\frac 1{z^p} =\frac {\pi^2\k^2}4\,V_p^{(1)}(\k)
\label{C2-A2V}
\ee

\subsection{The eigenvalue of $\tilde A$ and $C$}

We proceed as in section 5.5, although here life is easier. In fact, due
to the structure of $\A$, see (\ref{A,C}),
the analogue of eq.(\ref{eigenA=0}) is trivial.
Therefore we have simply to prove
\be
\sum_{n=1}^\infty  A_{p,n} \,V_{n}^{(1)}(\k) ={\mathfrak a}(\k) \,
V_{p}^{(1)}(\k) \label{eigenAtw}
\ee
and find ${\mathfrak a}(\k)$. On the weight 1 basis $G\tilde\A -\tilde\A H^T=0$
reduces to $H^T\tilde A -\tilde A H^T=0$. Thanks to this and
(\ref{HTV1n}) we conclude that (\ref {eigenAtw}) is true. So we set out
to compute ${\mathfrak a}(\k)$.

Since $V_1^{(1)}(\k)=1$ we have
\be
\sum_{n=1}^\infty\, \tilde A_{1,n} V_n^{(1)}(\k) = {\mathfrak a}(\k) \,V_1^{(1)}(\k)
={\mathfrak a}(\k)
\label{alphatw}
\ee
On the other hand
\be
\tilde A_{1n} = (-1)^{\frac {n+1}2}\left(\frac 1n-\frac 1{n+2}\right)\0
\ee
Therefore
\be
{\mathfrak a}(\k) =- \sum_{l=0}^\infty (-1)^l
\left( \frac 1{2l+1}-\frac 1{2l+3}\right)\, V_{2l+1}^{(1)}(\k)=
G(1)-F(1)\label{twalpha1}
\ee
where
\be
&&F(z)= \sum_{l=0}^\infty (-1)^l \frac 1{2l+1} V_{2l+1}^{(1)}(\k) z^{2l+1}\label{twF}\\
&&G(z)= \sum_{l=0}^\infty (-1)^l \frac 1{2l+3} V_{2l+1}^{(1)}(\k) z^{2l+3}\label{twG}
\ee
So
\be
\frac {d F}{dz} &=& \sum_{l=0}^\infty (-1)^l V_{2l+1}^{(1)} z^{2l}=
\frac 12 \left(f_\k^{(1)}(iz) + f_\k^{(1)}(-iz)\right)\label{twdF}\\
\frac {d G}{dz} &=& \sum_{l=0}^\infty (-1)^l V_{2l+1}^{(1)} z^{2l+2}=
 \frac {z^2}2\,\left(f_\k^{(1)}(iz) + f_\k^{(1)}(-iz)\right)\label{twdG}
 \ee
Let us define
\be
H(z) = G(z)-F(z)\label{twH}
\ee
We have $H(0)=0$. So that ${\mathfrak a}(\k)=H(1)$.

Now
\be
\frac {d H}{dz} &=& -\cosh \left(\k\, \arctan (iz)\right)\0\\
&=&-\frac 12  \left( (1+z)^\zeta (1-z)^{-\zeta} +(\zeta\rightarrow -\zeta)\right)
\label{twdH}
\ee
where $\zeta = i\k/2$. Therefore
\be
H(1)&=&-\frac 12 \int_0^1 dz
\left((1+z)^\zeta (1-z)^{-\zeta}\, +\,(\zeta\rightarrow -\zeta)\right)\label{twH(1)}\\
&=& -\frac 1{2(1-\zeta)} \left(\, F(-\zeta,1;2-\zeta;-1)\,+\,
\zeta\rightarrow -\zeta\right)
\ee
Now we use (\ref{hyper4},\ref{hyper5}) and obtain
\be
H(1)=-\frac 1{2(1-\zeta)} \left( -\zeta\, F(1-\zeta,1;2-\zeta;-1)- (1-\zeta)\,
F(-\zeta,1;1-\zeta;-1)+(\zeta \rightarrow-\zeta)\right)\0
\ee
Next use (\ref{hyper3})
\be
H(1) = \frac {\zeta}4\, [  \EG(-\zeta) -\EG(1-\zeta) -\EG(\zeta)+\EG(1+\zeta)]=
\frac{\zeta}4\,\left(- \frac {4\pi}{\sin (\pi\zeta)} \right)\label{twH3}
\ee
i.e.
\be
{\mathfrak a}(\k)=\frac {\pi\k}
{2\,\sinh \left(\frac {\pi \k}2\right)}\label{ghalpha}
\ee

From the above results for $C^2-A^2$ and $A$ we can determine the eigenvalue
of $C$ up to a sign
\be
{\mathfrak c}(\k)=\pm \frac{\pi \k}2 \coth( \frac{\pi \k}2)\label{gammak}
\ee
That the correct sign is actually + is shown in Appendix E.

\subsection{Wedge states for twisted ghosts}

We have shown above that the solution to the KP equation in the twisted case
reduces to the small matrix $\alpha(t)$. Thanks to the relation
(\ref{relACMatTwgh}) with the matter we can now conclude that
\be
\alpha(t) = E\,\alpha^{(m)}(t)\,E^{-1} \label{refalphamattwgh}
\ee
Therefore we can dispense with the full discussion of the wedge states,
because we can use the results of the matter sector. We must simply bear
in mind the sign difference between the ghost and matter sector.
So, for instance,
\be
X^{(tw)} = -E X^{(m)} E^{-1}, \quad\quad T^{(tw)} =- E T^{(m)} E^{-1}
\ee
where the LHS matrices refer to the twisted ghosts. Moreover
the scalar function $\eta$ is the same for matter and twisted ghosts.
We can conclude that the equivalence (\ref{expo}) holds for twisted
ghost sector as well.

In this derivation we do not need to explicitly use the weight 0 basis, but
of course, such a basis is needed in order for the formalism to work.
A short discussion is given in Appendix E.

\section{The conventional n.o.}

In this last part of the paper we discuss the
conventional normal ordering introduced in section 3.1 and its problems.
Eventually it
will be clear that the treatment with this normal ordering is
very awkward, to say the least. However we cannot completely exclude
that the formalism may work with this normal ordering either.

\subsection{Commuting matrices}

Let us return to the matrices of subsection 3.1.
Using the formulas of the subsection 3.4 we can easily show
some properties that may help us integrating the KP equations.

Suppose $p$ and $\bar q$ are even. Then
\be
(A D^T)_{p \bar q} &=& \sum_{\bar k=2}^\infty A_{p\bar k}\,
D^T_{\bar k \bar q}=
\sum_{l=1}^\infty (2p+2l)(4l-\bar q) \ell_{|\bar q-2 l|} \,
\ell_{2l+p}\0\\
&=& 8 (-1)^{\frac {p+\bar q}2} \sum_{l=1}^\infty \frac {(p+l)(4l -\bar q)}
{((p+2l)^2-1)(((2l-\bar q)^2-1)}\label{ADTpqee}
\ee
having set $\bar k= 2l$. We notice that the RHS of this equation coincide with
(\ref{ACpqeven}). Similarly
\be
&&(C A)_{p \bar q} = \sum_{k=0}^\infty C_{p k}\,
A_{k \bar q}=
\sum_{l=0}^\infty (2p-2l)(4l+\bar q) \ell_{p+2 l} \,
\ell_{|2l-\bar q|}\0\\
&=& 8 (-1)^{\frac {p+\bar q}2}\left[ \sum_{l=1}^\infty \frac {(p-l)(4l +\bar q)}
{((p-2l)^2-1)(((2l+\bar q)^2-1)} + \frac {p\bar q}{(p^2-1)(\bar q^2-1)}\right]
\label{CApqee}
\ee
The first term in the square bracket on the RHS of this equation is
nothing but eq.(\ref{CApqeven}).
Putting everything together we find
\be
(A D^T - C A)_{p\bar q} = 8 (-1)^{\frac {p+\bar q}2}
\left[\frac {p\bar q}{(p^2-1)(\bar q^2-1)}-
\frac {p\bar q}{(p^2-1)(\bar q^2-1)} \right]=0\label{ADT-CA}
\ee
This derivation becomes singular in the case $p=0,\bar q=2$. This case has to
be dealt with separately. One easily gets
\be
(A D^T)_{02}= -\frac 14, \quad\quad (C A)_{02} =-\frac 14\label{02case}
\ee

Similarly for $p$ and $\bar q$ odd we have
\be
(A D^T)_{p \bar q} &=& \sum_{\bar k=1}^\infty A_{p\bar k}\,
D^T_{\bar k \bar q}=4 (-1)^{\frac {p+\bar q}2}
\sum_{l=0}^\infty \frac {(2p+2l+1) (4l -\bar q +2)}
{((p+2l+1)^2-1)(((2l-\bar q+1)^2-1)} \label{ADTpqoo}
\ee
and
\be
(C A)_{p \bar q} &=& \sum_{\bar k=0}^\infty C_{p k}\,
A_{k \bar q}= 4(-1)^{\frac {p+\bar q}2}
\sum_{l=0}^\infty \frac {(2p-2l-1)(4l +\bar q+2)}
{((p-2l-1)^2-1)(((2l+\bar q+1)^2-1)}\label{CApqoo}
\ee
having set $k = 2l+1$. The first term in square brackets in (\ref{ADTpqoo})
coincides with eq.(\ref{ACpq}), while the corresponding term in
the RHS of (\ref{CApqoo}) coincides with (\ref{CApq}). Therefore they are
equal. It follows that the RHS of (\ref{ADTpqoo}) equals the RHS of
(\ref{CApqoo}). The previous analysis does not hold for $p=1$ and $q=1$.
By a direct calculation we find
\be
(AD^T-CA)_{1,1} = [\CA,\CC]_{1,1}= \frac {\pi^2}8 \label{wrong11}
\ee
on the basis of (\ref{AC11}). This is the first obstacle brought in by the
conventional n.o.. One could adopt the attitude that this is a very marginal
non--commutativity and adopt a prescription in order to eliminate it.
One possibility in this sense is to state that
$(AD^T-CA)_{p,q}$ should be calculated for generic complex $p,q$ and then
analytically continued to integral $p,q$. The result is 0 and the
above discrepancy is eliminated. Let us assume this attitude and continue the
analysis.

Therefore we can conclude that
\be
A D^T = C A \label{ADT=CA}
\ee
which is the appropriate `commutation relation' for these kind of matrices.

In the same way one can prove that
\be
BC = D^T B\label{BC=DTB}
\ee

It follows again from these results that, for instance,
\be
C A B = A D^T B = A B C \label{CAB}
\ee

Another important identity is the one concerning
$C^2 -AB$. For odd $p,q$ we get
\be
(C^2-AB)_{pq}&=& \sum_{k=1} C_{pk} C_{kp}-\sum_{k=1} A_{pk} B_{kp}=
(\CC^2-\CA^2)_{pq}\0\\
&=& \frac{\pi^2}2 \left( (p^2-2)\,\delta_{p,q}+ \frac 12 p(p-1)
\,\delta_{q,p+2}
+\frac 12 p(p+1) \,\delta_{p,q+2}\right)\label{C2-A2oddconv}
\ee
and for even $p,q$
\be
(C^2-AB)_{pq}&=& \sum_{k=0} C_{pk} C_{kp}-\sum_{k=1} A_{pk} B_{kp}=
(\CC^2-\CA^2)_{pq} + C_{p0} C_{0q}\0\\
&=& (\CC^2-\CA^2)_{pq}
- 8 (-1)^{\frac {p+q}2} \frac {pq}{(p^2-1) (q^2-1)}\0\\
&=&\frac{\pi^2}2 \left( (p^2-2)\delta_{p,q}+ \frac 12 p(p-1) \,\delta_{q,p+2}
+\frac 12 p(p+1) \,\delta_{p,q+2}\right)\label{C2-A2evenconv}
\ee
These two are identities valid for long--legged square matrices.
We can prove in a similar way
\be
((D)^T-BA)_{\bar p \bar q}= \frac{\pi^2}2 \left( (\bar p^2-2)
\delta_{\bar p,\bar q}+ \frac 12 \bar p(\bar p-1) \,\delta_{\bar q,\bar p +2}
+\frac 12 \bar p (\bar p+1) \,\delta_{\bar p,\bar q+2}\right)
\label{C2-A2conv}
\ee
for square short--legged matrices.

\subsection{Integration of the KP equations}

The relevant KP equations
\be
\dot {\alpha}=A +C {\alpha} +{\alpha} D^T +{\alpha}  B
{\alpha}\label{KPconv}\\
\ee
and
\be
\dot \eta = -{\rm Tr}\left(\alpha\, B\right)\label{etaconv}
\ee
can be integrated as in section 4.2.
Proceeding in the same way it is easy to find the two solutions
\be
\alpha_1(t) = \frac {{\rm sinh}\left(\sqrt{C^2-AB}\,t\right)}
{ \sqrt{C^2-AB}\, {\rm cosh} \left(\sqrt{C^2-AB}\,t\right)
-C\, {\rm sinh}\left(\sqrt{C^2-AB}\,t\right)}\,A\label{1stalphatconv}
\ee
and
\be
\alpha_2(t) = A \frac {{\rm sinh}\left(\sqrt{(D^T)^2-BA}\,t\right)}
{ \sqrt{(D^T)^2-BA}\, {\rm cosh} \left(\sqrt{(D^T)^2-BA}\,t\right)
-D^T\, {\rm sinh}\left(\sqrt{(D^T)^2-BA}\,t\right)}
\label{2ndalphatconv}
\ee
These two solutions have the same functional form as in subsection 4.2,
but, of course, they have different legs. If the commutation rules
of the previous subsection hold (for that we need to adopt the analytic
continuation argument), we can use the same argument as in subsection 4.2
and conclude that these two solutions coincide.

\subsection{Other difficulties}

In the ghost sector with the conventional normal ordering we get
\be
K_1= \sum_{p,q\geq 0} c_p^\dagger\, G_{pq} \,b_q + \sum_{p,q\geq 1}
b_{\bar p}^\dagger\,H_{\bar p\bar q}\,
c_{\bar q}-c_1\,b_0 +c_0^\dagger b_1^\dagger\label{K1ghconv}
\ee
where the expression $G,H$ are the same as in (\ref{GF})
and $G$ is a is a square long--legged matrix and $H$ a square
short--legged one, but with a different length with respect to section 5:
long legs run from 0 to $+\infty$, short legs from 1 to $+\infty$.
This is not without consequences. Absorbing the last two terms on the RHS
of (\ref{K1ghconv}) in the first two, as we have done in the natural n.o.
case, requires stretching both legs of $G$ and $H$, and this complicates
a lot the task of finding simple commutation rules with $A, B,C,D$.
Trying then to diagonalize the matrices involved we face similar problems.
The matrix $A$ for instance is right--short--legged, but not as short as
the weight 2 basis, which starts at $n=2$. We do not exclude that all these
discrepancies might somehow be fixed. But at the moment we do not see how to
keep under control the consequences of the new prescriptions that are needed.
For this reason we leave the use of the conventional n.o. as an open problem.

\section{Conclusions}

Our precise aim in this paper was to prove the equivalence (\ref{nL0}) in the
oscillator formalism, in the matter and in the ghost sector separately.
And, starting from the RHS of (\ref{nL0}) we have succeeded in proving
the recursion relations characteristic of the diagonalized wedge states
matrices that feature in the LHS. This result is of course important for the
proof of the Schnabl's solution in the oscillator language, but
it is interesting in itself because wedge states are a powerful approximation
tool in SFT.

Some aspects of our proof should be underlined.
Although at first the task seems to be unwieldy in the ghost sector, because the
latter is characterized by asymmetric bases, it is nevertheless possible to
do it because the relevant infinite matrices can nevertheless be diagonalized.
We have learned from our research that such problems can be dealt with
in a way not devoid of elegance. We have also learned that the most suitable choice
for normal ordering in the ghost sector is the natural one. This of course
means that in future developments
(in particular in the continuation of the present program)
we have to reformulate the star product, the wedge states and so on in the
natural n.o.\footnote{We have already mentioned that strictly speaking,
eq.(\ref{nL0}) is not really demonstrated before this is done.}, while so far
they have been dealt with essentially with the conventional
n.o.. On the other hand the natural normal ordering seems to be the `natural'
one in the Schnabl's proof, otherwise we cannot see how one can deal with
such states as $c_1|0\rangle$.

\acknowledgments

L.B. would like to thank G.~F.~Dell'Antonio and M.~Schnabl for discussions
and in particular I.Arefe'eva for making him aware of ref.\cite{BeLove}.
L.B. would like to thank the CBPF (Rio de Janeiro) and the GGI (Florence)
for their kind hospitality and support during this research.
R.J.S.S would like to thank CBPF (Rio de Janeiro) for the kind
hospitality during this research.

This research was supported for L.B. by the Italian MIUR
under the program ``Superstringhe, Brane e Interazioni Fondamentali''.

C.M. was supported in part by the Belgian Federal Science Policy
Office through the Interuniversity Attraction Pole P5/27, in part
by the European Commission FP6 RTN programme MRTN-CT-2004-005104
and in part by the
``FWO-Vlaanderen'' through project G.0428.06.

D.D.T. was supported by the Science Research Center Program
of the Korea Science and Engineering Foundation through the Center for
Quantum Spacetime(CQUeST) of Sogang University with grant number
R11 - 2005 - 021..

\section*{Appendix}
\appendix

\section{Some remarkable identities}

In this section we present an explicit sample of calculations
needed in this paper.

We are going to repeatedly use the identity
\be
\frac{l}{(p+2l+a)(q+2l+b)}= \frac 1{2(p-q+a-b)}\left(\frac{p+a}{p+2l+a}-
\frac{q+b}{q+2l+b}\right)\label{basicid}
\ee
Now assume $p,q$ are both even.
\be
&&A^2_{pq}=2\sqrt{pq} \sum_{l=1}^\infty\, l
\frac {(-1)^{p+q}}{(p+2l)^2-1)((q+2l)^2-1)}\0\\
&&= \frac {\sqrt{pq}(-1)^{p+q}}2 \sum_{l=1}^\infty\, l
\left( \frac 1{p+2l-1}-\frac 1{p+2l+1}
\right) \left(\frac 1{q+2l-1}-\frac 1{q+2l+1}\right)\0\\
&&=\frac {\sqrt{pq}(-1)^{p+q}}2 \sum_{l=1}^\infty\,
\left[\frac 1{2(p-q)}\left(\frac {p-1}
{p+2l-1}- \frac {q-1}{q+2l-1}\right)\right. \0\\
&&~~~~~~~+\frac 1{2(p-q)}\left(\frac {p+1}
{p+2l+1}- \frac {q+1}{q+2l+1}\right)\0\\
&&~~~~~~~- \frac 1{2(p-q+2)}\left(\frac {p+1}
{p+2l+1}- \frac {q-1}{q+2l-1}\right) \0\\
&&~~~~~~~\left. -\frac 1{2(p-q-2)}\left(\frac {p-1}
{p+2l-1}- \frac {q+1}{q+2l+1}\right)\right] \label{Asq}
\ee
where use have been made of (\ref{basicid}).
Now we use the definition of the $\psi$ function as an infinite series
\be
\psi(z)= -\gamma-\frac 1z +\sum_{n=1}^\infty \frac z{n(z+n)}\0
\ee
which can be written, using $\frac z{n(z+n)}=\frac 1n -\frac 1{z+n}$,
\be
\sum_{n=1}^\infty \frac 1{z+n} = -{\rm log} \epsilon -
\left(\frac 1z +\gamma+ \psi(z)\right)\label{defpsi}
\ee
where $\epsilon$  is a regulator for $\epsilon\to 0$:
$-{\rm log} \epsilon = \sum_{n=1}^\infty (1-\epsilon)^n/n$. Of course anything
meaningful must not depend on $\epsilon$.

Replacing (\ref{defpsi}) inside (\ref{Asq}) one finds that all terms cancel out
except those containing $\psi$, so that one gets
\be
&&A^2_{pq}= \frac{\sqrt{pq} (-1)^{p+q}}8\left[ - \left(\frac{p-1}{p-q}
-\frac{p-1}{p-q-2}\right) \psi\left(\frac {p-1}2\right)
+ \left(\frac{q-1}{p-q} -\frac{q-1}{p-q+2}\right)\psi\left(\frac{q-1}2\right)\right.\0\\
&&\left. - \left(\frac{p+1}{p-q} -\frac{p+1}{p-q+2}\right)\psi\left(\frac {p+1}2\right)
+\left(\frac{q+1}{p-q} -\frac{q+1}{p-q-2}\right)\psi\left(\frac {q+1}2\right)\right]\0\\
&=& \frac{\sqrt{pq} (-1)^{p+q}}4\left[ \frac{p-1}{p-q-2}\psi\left(\frac {p-1}2\right)+
\frac{q-1}{p-q+2}\psi\left(\frac {q-1}2\right)\right.\0\\
&&- \left.\frac{p+1}{p-q+2} \psi\left(\frac {p+1}2\right)
-\frac{p+1}{p-q-2}\psi\left(\frac {q+1}2\right)\right]\0\\
&=&\frac{\sqrt{pq} (-1)^{p+q}}{2(p-q)} \left[ \frac{p+q}{(p-q)^2-4}
\psi\left(\frac {p+1}2\right)- \frac{p+q}{(p-q)^2-4}
\psi\left(\frac {q+1}2\right)- \frac  {2(p-q)}{(p-q)^2-4}\right]\label{Asquare}
\ee

For $C^2$ one proceeds in the same way
\be
&&C^2_{pq}= 8\sqrt{pq} \sum_{l=1}^\infty\, l \frac {(-1)^{p+q}}
{(p-2l)^2-1)((q-2l)^2-1)}\0\\
&=& -\frac{\sqrt{pq} (-1)^{p+q}}2\left[ \left(\frac{p-1}{p-q}
-\frac{p-1}{p-q-2}\right) \psi\left(\frac {1+p}2\right)
- \left(\frac{q-1}{p-q} -\frac{q-1}{p-q+2}\right)\psi
\left(\frac{1+q}2\right)\right.\0\\
&&\left. + \left(\frac{p+1}{p-q} -
\frac{p+1}{p-q+2}\right)\psi\left(\frac {3+p}2\right)
-\left(\frac{q+1}{p-q} -
\frac{q+1}{p-q-2}\right)\psi\left(\frac {3+q}2\right)\right]\0
\ee
where use has been made of
\be
\psi\left(\frac 12 +z\right) =
\psi\left(\frac 12 -z\right)+\pi \, {\rm tan}(\pi\,z)\label{psitan}
\ee
This can be reduced to
\be
C^2_{pq}= \frac{2\sqrt{pq} (-1)^{p+q}}{(p-q)} \left[ \frac{p+q}{(p-q)^2-4}
\psi\left(\frac {1-p}2\right)- \frac{p+q}{(p-q)^2-4}
\psi\left(\frac {1-q}2\right)- \frac  {2(p-q)}{(p-q)^2-4}\right]\0
\ee
Therefore $C^2_{pq}-4 A^2_{pq}$ is the same as in eq.(\ref{C2-4A2}).

The $p,q$  both odd case can be dealt with in a similar way.

\section{The weight 2 and -1 bases}.

A series of bases of weight $s$ were introduced in \cite{Belov1,Belov2}.
They are
expressed in terms of generating functionals $f^{(s)}_{\k}(z)$, where $s$ is
integer or half--integer. In order to normalize them
the quadratic form
\be
\langle f |g\rangle = \frac 1{\pi \,\Gamma(2s-1)} \int_{|z|\leq 1}
d^2z \, (1-|z|^2)^{2s-2} \overline {g(z)} \,f(z)\label{quadrform}
\ee
is used.
This leads to the normalized generating functions
\be
\tilde f^{(s)}_{\k}(z)= \sqrt{A_s(\k)} f^{(s)}_{\k}(z)\label{nomgenfun}
\ee
where $f^{(s)}_{\k}(z)$ are the generating functions we have used in the text
(eigenfunctions of ${\cal K}_1$) and
\be
A_s(\k)= \frac{2^{2s-2}}{\pi}\, \Gamma\left( s+\frac {i\k}2\right)\,
\Gamma\left( s-\frac {i\k}2\right),\label{Ask}
\ee
which satisfies
\be
A_{s+1}(\k) = (\k^2+4s^2) A_s(\k) \0
\ee

In \cite{Belov1,Belov2}
\be
\tilde f^{(s)}_{\k}(z)\equiv |\k,s\rangle (z) \equiv \langle z,s|\k,s\rangle
= \overline{\langle \k,s|z,s\rangle}\label{fzk}
\ee
where we have introduced two continuous basis,
the $\k$ basis and the $z$ basis, \cite{Oku2}. The normalization (\ref{Ask}) is such
that
\be
\langle \k,s|\k',s\rangle = \delta(\k,\k') \label{deltakk'}
\ee

Let us also introduce the discrete basis
$|n,s\rangle$, such that
\be
\langle n,s|z,s\rangle = N_n^{(s)} z^{n-s}, \quad\quad n\geq s
\label{discreteb}
\ee
The discrete basis satisfies $\langle n,s|m,s\rangle=\delta_{n,m}$, therefore
it is the basis of the space of square--summable sequences.
Normalizing according to (\ref{quadrform}) one gets
\be
N_n^{(s)} = \sqrt{\frac {\Gamma(n+s)}{\Gamma(n-s+1)}}\label{Nns}
\ee
The polynomials (\ref{discreteb}) form a complete set of orthonormalized
eigenfunctions of $L_0$.
We notice that the normalization constant $N_n^{(s)}$
are singular corresponding to the ghost zero modes (i.e. for $n=0$, $s=0$
and $n=-1,0,1$ for $s=-1$). In these cases we can use  $s$ as a regulator
\be
&&N_0^{(s)} = \frac 1{\sqrt{s}}, \quad\quad s\approx 0\0\\
&& N_1^{(s)} = N_{-1}^{(s)} = \frac 1{\sqrt{2(s+1)}}, \quad\quad
N_0^{(s)}=\frac 1{\sqrt{-1-s}}, \quad\quad s\approx -1\label{N-101}
\ee

Inserting a discrete basis in (\ref{fzk})
we get
\be
\tilde f^{(s)}_{\k}(z)=\langle z,s|\k,s\rangle =
\sum_{n=s}\langle z,s|n,s\rangle\langle n,s|\k,s\rangle
= \sum_{n=s} N_n^{(s)} \langle n,s|\k,s\rangle z^{n-s} \label{fzk2}
\ee
Therefore
\be
\tilde V_n^{(s)} (\k) = N_n^{(s)} \,  \langle n,s|\k,s\rangle\label{fzk3}
\ee
Now, after defining
\be
\hat V_n^{(s)} (\k)= \langle n,s|\k,s\rangle=
\frac {\sqrt{A_s(\k)}}{N_n^{(s)}} \, V_n^{(s)}(\k)\label{Vnshat}
\ee
one can prove the orthonormality relation
\be
\int_{-\infty}^{\infty} d\k\, \hat V_n^{(s)} (\k)\hat V_m^{(s)} (\k)=
\delta_{n,m}\label{orthVnVm}
\ee
And introducing a complete set of discrete states in (\ref{deltakk'})
one gets the completeness relation
\be
\sum_{n=s} \hat V_n^{(s)} (\k)\hat V_n^{(s)} (\k')=\delta(\k,\k')\label{completeness}
\ee

In conclusion, all the spaces ${\cal H}^{(s)}$ possess a complete
orthonormal basis
given by $\hat V_n^{(s)}$. However {\it this is not a basis of eigenfunctions
of the matrices $G$ or $H$}. The eigenfunctions of $G$ or $H$ are given
by the $\tilde V_n^{(s)}$ or by the corresponding unnormalized ones,
$V_n^{(s)}$. What is relevant for the latter is a {\it biorthogonality}
relation.

Let us concentrate on the conjugate cases $s=-1$ and $s=2$, that is
the weight -1 and weight 2 basis of section 3. In an
unpublished paper, using the same techniques as in \cite{Belov1},
Belov and Lovelace, \cite{BeLove}, showed that
\be
\int_{-\infty}^{\infty} d\k\, \tilde V^{(-1)}_{n}(\k)\,\tilde
V^{(2)}_{m}(\k)= \delta_{n,m}, \quad\quad n\geq 2\label{biorthog}
\ee
that is the two basis are biorthogonal. Moreover we have the completeness
relation
\be
\sum_{n=2}^\infty \, \tilde V^{(-1)}_{n}(\k)\, \tilde V^{(2)}_{n}(\k')=
\delta(\k,\k')\label{bi-completeness}
\ee

We would like now to give an example of application of these
bases which underlies our approach in this paper. Let us consider
$c^\dagger A b^\dagger$. It can be written as follows
\be
c^\dagger A b^\dagger &=& \sum_{n=-1, m=2} c^\dagger_n A_{nm} b_m^\dagger
=  \sum_{n=-1, m=2}^\infty \int d\k\, d\k'\,\tilde c(\k) \tilde V_n^{(-1)}(\k)
\tilde A_{nm}  \tilde V_m^{(2)}(\k')\, b(\k')\0\\
&=& \sum_{n=2}^\infty \int d\k\, d\k'\, \tilde c(\k)\tilde V_n^{(-1)}(\k)
\tilde A(\k) \tilde V_n^{(2)}(\k')b(\k')=
\int d\k\,\tilde c(\k)\tilde A(\k)\, b(\k)\label{cABdagger}
\ee
where we have introduced
\be
(-1)^n\, c_n^\dagger = \int d\k \,\tilde c(\k) \,V_n^{(-1)}(\k),\quad\quad
b_n^\dagger = \int d\k\, b(\k)\, V_n^{(2)}(\k)\label{c(k)}
\ee
and used (\ref{biorthog}).

We can reverse (\ref{c(k)}) again by means of (\ref{biorthog})
\be
\tilde c(\k) = \sum_{n=2}^\infty (-1)^n c_n^\dagger \, V_n^{(2)}(\k) ,
\quad\quad b(\k) = \sum_{n=2}^\infty  b_n^\dagger \, V_n^{(-1)}(\k)
\ee

We note that in (\ref{cABdagger}) the modes $c_a$ with $a=-1,0,1$ are
irrelevant. Moreover the eigenfunctions corresponding to these three zero
modes are excluded from the completeness relation (\ref{bi-completeness}).
We could qualitatively phrase the reason for that
by saying that they do not carry more
information about the system than what is already contained in the
remaining eigenvectors. They can in fact be expressed in terms of the latter
due to the relations
\be
&& \int_{-\infty}^{\infty} d\k\, \tilde V^{(-1)}_{-1}(\k)\, \tilde f^{(2)}_{\k}(z)
= \frac z{(1+z^2)^2} \label{V-1-1}\\
 &&\int_{-\infty}^{\infty} d\k\, \tilde V^{(-1)}_{0}(\k)\, \tilde f^{(2)}_{\k}(z)
= \frac 1{(1+z^2)} \label{V-10}\\
&&\int_{-\infty}^{\infty} d\k\, \tilde V^{(-1)}_{1}(\k)\,  \tilde f^{(2)}_{\k}(z)
= \frac z{(1+z^2)}+\frac z{(1+z^2)^2} \label{V-11}
\ee
For let us write $\tilde V_a^{(-1)}(\k) = \sum_{n=2}^\infty \, b_{an}
\tilde V_n^{(-1)}(\k)$, for $a=-1,0,1$ and plug these expressions
in the LHS of (\ref{V-1-1},\ref{V-10},\ref{V-11}).  By expanding
both sides of the equations in powers of $z$ and equating coefficients
of the same powers we can determine all the $b_{an}$ coefficients
(they are the same as the coefficients of the power series expansion
in $z$ in rhs of (\ref{V-1-1},\ref{V-10},\ref{V-11}).
Therefore the $\tilde V_a^{(-1)}(\k)$ do not contain additional information
with respect to the set of  $\tilde V_n^{(-1)}(\k)$ with $n\geq 2$.

It would seem that, using this result, we can come to the absurd conclusion
that $\hat V_a^{(-1)}(\k)$ can be expanded in terms of
$\hat V_n^{(-1)}(\k)$ with $n\geq 2$. However the true relation is
\be
\hat V_a^{(-1)}(\k) = \sum_{n=2}^\infty \, b_{an} \frac{N_n^{(-1)}}{N_a^{(-1)}}
\hat V_n^{(-1)}(\k)\label{absurd}
\ee
But  due to the properties (\ref{N-101}) this equation means nothing but
$0=0$.

\section{Properties of the hypergeometric functions}

In this section we collect some properties of the hypergeometic function
$F(a,b;c,z)\equiv {}_2 \! F_1(a,b;c;z)$  and other special functions
that we need in various derivations
of this paper. For their derivation, see for instance \cite{Erdelyi}.
We start with the integral representation of the
hypergeometric function
\be
F(a,b;c;z)=\frac {\Gamma(c)}{\Gamma(b)\Gamma(c-b)}
\int_0^1 dt\, t^{b-1}(1-t)^{c-b-1}(1-tz)^{-a}\label{hyper1}
\ee
which is valid when $Re(c)>Re(b)>0$ and $|Arg(1-z)|<\pi$.

Next we need some identities valid for the special case of the argument
$z=-1$:
\be
F(a,b;1+a-b;-1)= 2^{-a} \frac {\Gamma(1+a-b) \,\sqrt{\pi}}{\Gamma(1-b+\frac 12 a)\,
\Gamma(\frac 12+\frac a 2)}\label{hyper2}
\ee
\be
F(a,b;b,-1) = 2^{-a}\label{hyper0}
\ee
and
\be
F(a,1;a+1;-1) = \frac a2\,\left(\psi\left(\frac 12+\frac a 2\right)-
\psi\left(\frac a2\right)\right)\equiv  \frac a2\,\EG(a)
\label{hyper3}
\ee
where $\psi$ is the dilogarithm function and $\EG$ is another
special function defined by (\ref{hyper3}).

Next we need other raising and lowering parameters relations:
\be
F=-\frac 1{c-a-1} \left(a\,F(a+1) -(c-1) F(c-1)\right)\label{hyper4}
\ee
and
\be
F= - \frac 1{c-a-b} \left(a(1-z)\,F(a+1)-(c-b)F(b-1)\right)\label{hyper5}
\ee
In these formulas the arguments are not indicated if they are the obvious
ones $a,b,c,z$.

In order to raise $b$ one can use
\be
&&F(a,b;c;-1) = -\frac {3+3b-c-a}{c-b-1} F(a,b+1;c;-1) +
2\frac {b+1}{c-b-1} F(a,b+2;c;-1)\label{hyper6}\\
&&F(a,b;c;-1) = \frac {3c-b-a+1}{2c} F(a,b;c+1;-1) +
\frac {(c-b+1)(c-a+1)}{2c(c+1)} F(a,b;c+2;-1)\0\\
\label{hyper7}
\ee

\section{The eigenvalues of $\tilde A$ and $D^T$}

\subsection{Zero modes of $\EA$}

Here we would like to prove eq.(\ref{eigenA=0}). Let us compute
\be
\sum_{n=2}^\infty A_{-1,n} V_n^{(2)}(\k)
&=&\frac 1{2\pi i}\oint dz\,\frac {e^{\k \arctan(z)}}{(1+z^2)^2}
\sum_{l=1}^\infty \frac {2 (-1)^{l}}{(2l+1)z^{2l}}\0\\
&=&\frac 1{2\pi i}
\oint dz\,\frac {e^{\k \arctan(z)}}{(1+z^2)^2}
\left(-2 +2z\, \arctan \left( \frac 1z \right)\right) \label{A-1V}
\ee
Similarly
\be
\sum_{n=2}^\infty A_{0,n} V_n^{(2)}(\k)&=&\frac 1{2\pi i}
\left(z +(1-z^2)\, \arctan \left( \frac 1z \right)\right)\label{A0V}\\
\sum_{n=2}^\infty A_{1,n} V_n^{(2)}(\k)&=&\frac 1{2\pi i}
\oint dz\,\frac {e^{\k \arctan(z)}}{(1+z^2)^2}
\left(-2 +2z\, \arctan \left( \frac 1z \right)\right)\label{A1V}
\ee
Now we use
\be
\arctan \left( \frac 1z \right)=
\frac i2 \log \left(\frac {iz+1}{iz-1}\right) =
-\arctan (z) \pm \frac {\pi}2\label{discequat}
\ee
It is evident that inserting (\ref{discequat}) in the RHS's of
(\ref{A-1V},\ref{A0V},\ref{A1V}) we get 0, because there is no pole
left at the origin. So eq.(\ref{eigenA=0}) is justified.
The function $ \arctan \left( \frac 1z \right)$ has two branch point at
$\pm i$, with a cut between them. There are two sheets and the sign
reflects the choice of which sheet we choose to do the integration on.
Both choices give a vanishing result. The point is that
$ \arctan \left( \frac 1z \right)$ converge for large $z$, but of course we
can continue it analytically to small $z$, where we realize that it does not
have poles at the origin, so the integrals vanish. In other words summing all
the powers $1/z^n$ eliminates these poles at $z=0$.

Let us also add that we have obtained the same vanishing result
(\ref{eigenA=0}) applying the same technique we used in calculating
the eigenvalues in section 5.5.

\subsection{The eigenvalue of $D^T$ in the weight 2 basis}

Using the same technique as in section5.5 we can calculate the eigenvalue
of $D^T$.
Since $V_2^{(2)}(\k)=1$ we have
\be
\sum_{n=2}^\infty\, D^T_{2,n} V_n^{(2)}(\k) = {\mathfrak c}(\k) \,V_2^{(2)}(\k)
={\mathfrak c}(\k)
\label{gammak2}
\ee
On the other hand
\be
D^T_{2,n} = -(-1)^{\frac {n}2}\left(\frac 3{n-1}-\frac 1{n-3}\right)\0
\ee
Therefore
\be
{\mathfrak c}(\k) = -\sum_{l=1}^\infty (-1)^l
\left( \frac 3{2l-1}-\frac 1{2l-3}\right)V_{2l}^{(2)}(\k)=
-3F(1)+G(1)\label{gammak21}
\ee
where
\be
&&F(z)= \sum_{l=1}^\infty (-1)^l \frac 1{2l-1} V_{2l}^{(2)}(\k)
z^{2l-1}\label{Fc2}\\
&&G(z)= \sum_{l=1}^\infty (-1)^l \frac 1{2l-3} V_{2l}^{(2)}(\k)
z^{2l-3}\label{Gc2}
\ee
So
\be
\frac {d F}{dz} &=& \sum_{l=1}^\infty (-1)^l V_{2l}^{(2)} z^{2l-2}=
-\frac 12 \left(f_\k^{(2)}(iz) + f_\k^{(2)}(-iz)\right)\label{dFc2}\\
\frac {d G}{dz} &=& \sum_{l=1}^\infty (-1)^l V_{2l}^{(2)} z^{2l-4}=
- \frac 1{2z^2}\,\left(f_\k^{(2)}(iz) + f_\k^{(2)}(-iz)\right)\label{dGc2}
 \ee
Let us define
\be
H(z) = G(z)-3F(z)\label{Hc2}
\ee
So that ${\mathfrak c}(\k)=H(1)$. Notice that near 0, $H(z)= \frac 1z+\ldots$.

Now
\be
\frac {d H}{dz} &=&\frac {3z^2-1}{z^2(1-z^2)^2}\,\cosh \left(\k\,
\arctan (iz)\right)\label{dHdzc2}
\ee
We have to eliminate the singularity at $z=0$. We use
that $V_3^{(2)}(\k)=\k$. So
\be
\sum_{n=2}^\infty\, D^T_{3,n} V_n^{(2)}(\k) = {\mathfrak c}(\k) \,V_3^{(2)}(\k)
=\k\,{\mathfrak c}(\k)
\0
\ee
On the other hand
\be
D^T_{3,2l+1} = -2(-1)^{l}\left(\frac 2{2l-1}-\frac 1{2l-3}\right)\0
\ee
Therefore
\be
\k{\mathfrak c}(\k) =- 2\sum_{l=1}^\infty (-1)^l
\left( \frac 2{2l-1}-\frac 1{2l-3}\right)V_{2l+1}^{(2)}(\k)=
2(G'(1)-2F'(1))\label{gammac2'}
\ee
where
\be
&&F'(z)= \sum_{l=1}^\infty (-1)^l \frac 1{2l-1} V_{2l+1}^{(2)}(\k) z^{2l-1}
\label{F'c2}\\
&&G'(z)= \sum_{l=1}^\infty (-1)^l \frac 1{2l-3} V_{2l+1}^{(2)}(\k)
z^{2l-3}\label{G'c2}
\ee
So
\be
\frac {d F'}{dz} &=& \sum_{l=1}^\infty (-1)^l V_{2l+1}^{(2)} z^{2l-2}=
\frac i{2z} \left(f_\k^{(2)}(iz) - f_\k^{(2)}(-iz)\right)\label{dF'c2}\\
\frac {d G'}{dz} &=& \sum_{l=1}^\infty (-1)^l V_{2l+1}^{(2)} z^{2l-4}=
\frac i{2z^3} \,\left(f_\k^{(2)}(iz) - f_\k^{(2)}(-iz)\right)\label{dG'c2}
 \ee
Let us define
\be
H'(z) =G'(z)-2F'(z)\label{H'c2}
\ee
We have ${\k}{\mathfrak c}(\k)=2H'(1)$. Near 0 we have $H'(z)=\frac {\k}z+\ldots$.
Therefore the combination $\k H-H'=\hat H$ vanishes at 0.

Now
\be
\frac {d H'}{dz} &=&\frac {i(1-2z^2)}{z^3(1-z^2)^2}\, \sinh \left(\k\,
\arctan (iz)\right)\0
\ee

We have
\be
\hat H(0)=0,\quad\quad \hat H(1) = \frac {\k}2 {\mathfrak c}(\k)\label{hatHc2}
\ee
So,
\be
\hat H(1) &=& \int_0^1 dz \left\{ \frac {3z^2-1}{2z^2} \frac \k{(1-z^2)^2}
\left[ \left(\frac {1+z}{1-z}\right)^{\frac {i\k}2}+
\left(\frac {1+z}{1-z}\right)^{-\frac {i\k}2}\right]\right.\0\\
&&~~~~~~-\left. \frac i{2\,z^3} \frac {1-2z^2}{(1-z^2)^2}
\left[ \left(\frac {1+z}{1-z}\right)^{\frac {i\k}2}-
\left(\frac {1+z}{1-z}\right)^{-\frac {i\k}2}\right]\right\}\label{hatHc2int}
\ee

Setting $\zeta=\frac {i\k}2$ and using the hypergeometric function $F={}_2F_1$
we can write formally
\be
\frac 1i\,\hat H(1)&=& \Gamma(-1-\zeta)\left[3\zeta\,
\frac {\Gamma(1)}{\Gamma(-\zeta)}\,F(2-\zeta,1;-\zeta;-1)\right.\0\\
&&-\zeta\,
\frac {\Gamma(-1)}{\Gamma(-2-\zeta)}\,
F(2-\zeta,-1;-2-\zeta;-1)-\frac {\Gamma(-2)}{\Gamma(-3-\zeta)}\,
F(2-\zeta,-2;-3-\zeta;-1)\0\\
&&\left.+2\frac {\Gamma(0)}{\Gamma(-1-\zeta)}\,F(2-\zeta,0;-1-\zeta;-1) -
(\zeta\rightarrow
-\zeta)\right]\label{hatHc2F}
\ee
which is evidently ill--defined.
The usual integral representation of the hypergeometric function (\ref{hyper1})
is valid when $Re(c)>Re(b)>0$ and $|Arg(1-z)|<\pi$. The last condition is
satisfied, but the former are not. We modify the $b$ parameter by shifting it
$b\to b+\beta$ so that (\ref{hatHc2F}) will become
\be
\frac 1i\,\hat H(1)&=& \Gamma(-1-\zeta)\left[3\zeta\,
\frac {\Gamma(1+\beta)}{\Gamma(\beta-\zeta)}\,F(2-\zeta,1+\beta;\beta-\zeta;-1)\right.\0\\
&&-\zeta\,\frac {\Gamma(-1+\beta)}{\Gamma(\beta-2-\zeta)}\,
F(2-\zeta,-1+\beta;\beta-2-\zeta;-1)\0\\
&&-\frac {\Gamma(-2+\beta)}{\Gamma(\beta-3-\zeta)}\,
F(2-\zeta,-2+\beta;\beta-3-\zeta;-1)\0\\
&&\left.+2\frac {\Gamma(\beta)}{\Gamma(\beta-1-\zeta)}\,
F(2-\zeta,\beta;\beta-1-\zeta;-1)
- (\zeta\rightarrow
-\zeta)\right]\label{hatHc2Freg}
\ee

Since we know the value of ${\mathfrak c}(\k)$ up to the sign from the
knowledge of $(D^T)^2-BA$ and ${\mathfrak a}(\k)$, we have only to check the sign.
Therefore it is enough to evaluate this expression numerically.
This can be done for instance with Mathematica. The limit
$\beta\to 0$ is a complicated function of $\zeta$, but coincides
exactly with ${\mathfrak c}(\k)$ we obtain from $((D^T)^2-AB)(\k)$ and
${\mathfrak a}(\k)$
for any value of $\k$, provided we choose the $+$ sign, i.e. we get
eq.(\ref{gamma}).
Of course it would be desirable to derive this result analytically.

\subsection{The eigenvalue of $\tilde A$ in the weight -1 basis}

It is important to verify that that the weight -1 basis left--diagonalizes
$\tilde A$
with the same eigenvalue as in the weight 2 basis. Here by eigenvalue
of $\tilde A$ we actually mean the eigenvalue of $\tilde \EA$. We expect
that the contribution of the 0 modes is irrelevant as in Appendix D1.
In that case we must have
\be
\sum_{n=-1}^\infty V_n^{(-1)}(\k)\tilde A_{n,2}={\mathfrak a}(\k)\,V_2^{(-1)}(\k)=
{1\over 6}\k(4+\k^2){\mathfrak a}(\k))
\label{eigeneqc}
\ee
as $V_2^{(-1)}(\k)={1\over 6}\k(4+\k^2)$. For $n= 2l$ we have
\be
\tilde A_{2l,2}=4(-1)^l \frac 1{2l+3}\label{A2nc}
\ee
Now define
\be
F(z)= \sum_{l=0}^\infty \frac{(-1)^l }{2l+3} V_{2l}^{(-1)}
(\k) z^{2l+3}\label{F(zc)}
\ee
so that ${\mathfrak a}(\k)\,V_2^{(-1)}(\k)=4F(1)$. On the other hand we have
\be
{dF\over dz}&=&\sum_{l=0}^\infty (-1)^l  V_{2l}^{(-1)} (\k) z^{2l+2}=
-\frac{iz}2\left(f_\k^{(-1)}(iz)-f_\k^{(-1)}(-iz)\right)\0\\
&=&-iz(1-z^2)\sinh(k\arctan(iz))
\ee
where $f_\k^{(-1)}$ is the weight -1 basis generating function.
Integrating this equation and noting $F(0)=0$ we obtain
\be
F(1)&=&-{i\over 2}\int_0^1 dz z\left( (1+z)^{1+\zeta}(1-z)^{1-\zeta}
-(1+z)^{1-\zeta}(1-z)^{1+\zeta}\right)\\
&=&-{i\over 2}\left({1\over (3-\zeta)(2-\zeta)} F(-1-\zeta,2;
4-\zeta;-1)\right. \0\\
&&\left. -{1\over (3+\zeta)(2+\zeta)}
F(-1+\zeta,2;4+\zeta;-1)\right)\0
\ee
where $\zeta={ik\over 2}$. Putting everything together we can write
\be
{\mathfrak a}(\k)=-{12i\over k(4+k^2)}\left({1\over (3-\zeta)(2-\zeta)}
F(-1-\zeta,2;4-\zeta;-1)-(\zeta\to -\zeta)\right)
\ee
We can check numerically that this result is exactly  the ${\mathfrak a}(\k)$
obtained from the b--basis. This can also be checked analytically by
repeatedly applying the equations in Appendix C. We begin by eq.(\ref{hyper6}),
which gives
\be
{\mathfrak a}(\k)=-{12i\over k(4+k^2)}\left({1\over 2(2-\zeta)}+
{\zeta\over (3-\zeta)(2-\zeta)} F(-1-\zeta,1;4-\zeta;-1)-
(\zeta\to -\zeta)\right)
\ee
where we have used $F(a,0;c;-1)=1$. Now we can apply
eq.(\ref{hyper7}) to obtain
\be
{\mathfrak a}(\k)&=&-{12i\over k(4+k^2)}\left[{1\over 2(2-\zeta)}+
{\zeta\over (2-\zeta)}\left(-{1\over 2}
F(-1-\zeta,1;2-\zeta;-1)\right.\right.\0\\
&+&\left.\left.{7-2\zeta\over 4(2-\zeta)}
F(-1-\zeta,1;3-\zeta;-1)\right)-(\zeta\to -\zeta)\right]
\ee
Repeating the same thing three times we will finally get
\be
{\mathfrak a}(\k)&=&-{12i\over k(4+k^2)}\left[-{\zeta\over (2-\zeta)}
\left({11-16\zeta + 2\zeta^2(7-2\zeta )\over 6}
F(-1-\zeta,1;-1-\zeta;-1)\right.\right.\0\\
&+&\left.\left.{2\zeta(2-\zeta)(1-\zeta)\over 3}
F(-1-\zeta,1;-\zeta;-1)+{1\over 2(2-\zeta)}\right)-(\zeta\to -\zeta)\right]\0
\ee
Using (\ref{hyper0}) and
(\ref{hyper3}) we can write this as
\be
{\mathfrak a}(\k)&=&-{12i\over k(4+k^2)}\left[{1\over 2(2-\zeta)}-
{\zeta\over (2-\zeta)}\left({11-16\zeta + 2\zeta^2(7-2\zeta )\over 12}\right.
\right.\0\\
&+&\left.\left.{2\zeta(2-\zeta)(1-\zeta^2)\over 6}
\left(\psi(-{1+\zeta\over 2})-\psi(-{\zeta\over 2})\right)\right)-
(\zeta\to -\zeta)\right].
\ee
Now we can proceed as in section 7.4 and put back $\zeta =\frac {i\k}2$ to get
\be
 {\mathfrak a}(\k)=\frac{\pi \k} 2  {\rm csch}({\pi k\over 2})
\ee

\section{The twisted ghost sector}

\subsection{The weight 0 basis}

Just as in the untwisted sector in the twisted one two bases are involved as
well. The first is the weight 1 basis we have introduced  in section 7.
The second basis corresponds to weight 0. For weight 0 ghost field $c$ we
have
\be
{\cal K}_1 \, c(z) =[K_1,c(z)]= (1+z^2)\partial c(z)\0
\ee
Integrating this equation we get that, if
\be
{\cal K}_1\, f_\k^{(0)}(z)= \k\, f_\k^{(0)}(z)\0
\ee
then
\be
f_\k^{(0)}(z)=  e^{\k\, {\arctan} (z)}=1+\k z+\ldots\label{fk0}
\ee
We set
\be
f_\k^{(0)}(z)=\sum_{n=0}^\infty \,V_n^{(0)}(\k) z^{n}\label{Vn0}
\ee
So, in particular, $V_0^{(0)}=1$, and
\be
V_n^{(0)}(\k)= \frac 1{2\pi i} \oint dz\,
\frac {e^{\k\,{\arctan}(z)}}{z^{n+1}}\label{Vn0oint}
\ee
So that, for instance, we can verify that
\be
\sum_{n=1} V_n^{(0)}(\k)\,G_{nq} &=&\frac 1{2\pi i} \oint dz\,
e^{\k{\,\arctan}(z)} n\, \sum_{n=0}^\infty (\delta_{q,n+1}+\delta_{q+1,n})
\,\frac 1{z^{n+1}}\0\\
&=&- \frac 1{2\pi i} \oint dz\,
e^{\k{\,\arctan}(z)}\frac d{dz} \left((1+z^2)\frac 1{z^{q+1}}\right)\0\\
&=& \k\, V_q^{(0)}(\k) \label{GV0n}
\ee
and, similarly,
\be
\sum_{n=0} H_{pn} V_n^{(0)}(\k) = \k\, V_p^{(0)}(\k)\0
\ee

\subsection{The eigenvalue of $C$}

The commutation rule $[H,\D]$=0, i.e $[H^T,\D^T]=0$, when applied to the
weight 1 basis, reduces to $[H^T,C]=0$. Therefore we conclude that
$C$ is diagonal in this basis. Since $V_1^{(1)}(\k)=1$ we have
\be
\sum_{n=1}^\infty\, C_{1,n} V_n^{(1)}(\k) = {\mathfrak c}(\k) \,V_1^{(1)}(\k)
={\mathfrak c}(\k)
\label{gammawc1}
\ee
On the other hand
\be
C_{1n} = (-1)^{\frac {n+1}2}\left(\frac 1{n-2}-\frac 1n\right)\0
\ee
Therefore
\be
{\mathfrak c}(\k) = -\sum_{l=0}^\infty (-1)^l
\left( \frac 1{2l-1}-\frac 1{2l+1}\right)V_{2l+1}^{(1)}(\k)=
-F(1)+G(1)\label{twgammac1}
\ee
where
\be
&&F(z)= \sum_{l=0}^\infty (-1)^l \frac 1{2l-1} V_{2l+1}^{(1)}(\k)
z^{2l-1}\label{twFc1}\\
&&G(z)= \sum_{l=0}^\infty (-1)^l \frac 1{2l+1} V_{2l+1}^{(1)}(\k)
z^{2l+1}\label{twGc1}
\ee
So
\be
\frac {d F}{dz} &=& \sum_{l=0}^\infty (-1)^l V_{2l+1}^{(1)} z^{2l-2}=
\frac{z^{-2}}2 \left(f_\k^{(1)}(iz) + f_\k^{(1)}(-iz)\right)\label{twdFc1}\\
\frac {d G}{dz} &=& \sum_{l=0}^\infty (-1)^l V_{2l+1}^{(1)} z^{2l}=
 \frac 12\,\left(f_\k^{(1)}(iz) + f_\k^{(1)}(-iz)\right)\label{twdGc1}
 \ee
Let us define
\be
H(z) = G(z)-F(z)\label{twH'}
\ee
So that ${\mathfrak c}(\k)=H(1)$. Notice that near 0, $H(z)= \frac 1z+\ldots$.

Now
\be
\frac {d H}{dz} &=&-{1\over z^2} \cosh \left(\k\,
\arctan (iz)\right)\label{dHdz}
\ee
If we try to integrate this function from $z=0$ to $z=1$
we will find a result involving $\Gamma(-1)$. Here we will
try to fix that.

Since $V_2^{(1)}(\k)=\k$ we have
\be
\sum_{n=1}^\infty\, C_{2,n} V_n^{(1)}(\k) = {\mathfrak c}(\k) \,V_2^{(1)}(\k)
=\k\,{\mathfrak c}(\k)
\label{gammaw2}
\ee
On the other hand
\be
C_{2,2l} = 2(-1)^{l}\left(\frac 1{2l-3}-\frac 1{2l-1}\right)\0
\ee
Therefore
\be
\k{\mathfrak c}(\k) = 2\sum_{l=1}^\infty (-1)^l
\left( \frac 1{2l-3}-\frac 1{2l-1}\right)V_{2l}^{(1)}(\k)=
2(F(1)-G(1))\label{twagamma}
\ee
where
\be
&&F(z)= \sum_{l=1}^\infty (-1)^l \frac 1{2l-3} V_{2l}^{(1)}(\k)
z^{2l-3}\label{twFc}\\
&&G(z)= \sum_{l=1}^\infty (-1)^l \frac 1{2l-1} V_{2l}^{(1)}(\k)
z^{2l-1}\label{twGc}
\ee
So
\be
\frac {d F}{dz} &=& \sum_{l=1}^\infty (-1)^l V_{2l}^{(1)} z^{2l-4}=
z^{-4}\frac{iz}2 \left(f_\k^{(1)}(iz) - f_\k^{(1)}(-iz)\right)\label{twdFc}\\
\frac {d G}{dz} &=& \sum_{l=1}^\infty (-1)^l V_{2l}^{(1)} z^{2l-2}=
 z^{-2}\frac{iz}2 \,\left(f_\k^{(1)}(iz) - f_\k^{(1)}(-iz)\right)\label{twdGc}
 \ee
Let us define
\be
H'(z) =2( F(z)-G(z))\label{twHc}
\ee
We have ${\k}{\mathfrak c}(\k)=H'(1)$. Near 0 we have $H'(z)=2\,\frac {\k}z+\ldots$.
Therefore the combination $2\k H-H'=\hat H$ vanishes at 0.

Now
\be
\frac {d H'}{dz} &=&2{i\over z^3} \sinh \left(\k\, \arctan (iz)\right)\0
\ee
Near $z=0$ this has the same behavior as $\frac {dH'}{dz}$ in (\ref{dHdz}).
We have
\be
\hat H(0)=0,\quad\quad \hat H(1) = \k{\mathfrak c}(\k)\label{hatH01}
\ee
Define
\be
{\mathfrak c}(\k)&=&\frac 1{\k}(2 \k H(1)-H'(1))\label{twghC(k)}\\
&=&\frac 1{\k}\int_{0}^{1}dz{1\over z^3}
\left[{\rm sin}\left({\k\over 2}\,{\rm Log}\left({1+z\over 1-z}\right)\right)
-\k z\,{\rm cos}\left({\k\over 2}\,{\rm Log}\left({1+z\over 1-z}\right)\right)\right]
\0
\ee
One can verify that the numerical values of ${\mathfrak c}(\k)$ at any
value of $\k$ coincides exactly with the ${\mathfrak c}(\k)$ in
(\ref{gammak}) and it selects the plus sign.

In the same way we can compute
\be
&&\sum_{n=0}^\infty V_n^{(0)}(\k)(C^2-A^2)_{nq} =\0\\
&&~~=\frac 1{2\pi i} \oint dz\, e^{\k{\,\arctan}(z)}
\cdot \frac {\pi^2}2 \sum_{n=0}^\infty\left( n^2\delta_{n,q} +\frac 14 n(n+q)
(\delta_{n,q+2}+\delta_{q,n+2})\frac 1{z^{n+1}}\right)\0\\
&&~~=\frac {\pi^2}4\,\frac 1{2\pi i} \oint dz\,
{e^{\k{\,\arctan}(z)}} \frac d{dz}\left( (1+z^2)\frac d{dz}\left((1+z^2)
 \frac 1{z^{q+1}}\right)\right)\0\\
& &~~=\frac {\pi^2\k^2}4\,V_q^{(0)}(\k)
\label{C2-A2V0}
\ee

\end{document}